\documentclass{iopjournal}
\usepackage{url}
\usepackage{mhchem} 
\usepackage{amsmath} 

\usepackage[style=numeric-comp,backend=biber,sorting=none,autocite=superscript]{biblatex}
\renewcommand{\cite}[1]{\supercite{#1}}
\usepackage{xcolor}
\usepackage{subcaption} 
\usepackage{longtable}
\usepackage{array} 
\usepackage{wrapfig} 
\usepackage{tablefootnote}

\begin{document}

\articletype{Paper} 

\title{Secondary Species formed from ionic liquid electrospray ion plume impacts with propellant thin films}

\author{Giuliana Caramella Hofheins$^{1, *}$\orcid{0000-0001-5575-2240}, Aleksandra B. Biedron$^2$\orcid{0000-0001-5403-7325} and Elaine M. Petro$^{1}$\orcid{0000-0002-4933-5405}}

\affil{$^1$Sibley School of Mechanical and Aerospace Engineering, Cornell University, Ithaca, USA}

\affil{$^2$Albany NanoTech Complex, Albany, USA}

\affil{$^*$Author to whom any correspondence should be addressed.}

\email{gch72@cornell.edu}


\begin{abstract}
The operational lifetime of ionic liquid electrospray propulsion systems is limited by plume–extractor electrode interactions. Over time, propellant accumulation, surface erosion, and electrical shorts degrade the extractor and therefore restrict the total impulse throughput. Characterizing the secondary species generated by plume impacts with deposited ionic liquid is therefore essential to understanding and mitigating these degradation pathways. A surface analysis technique known as Time-of-Flight Secondary Ion Mass Spectrometry (TOF-SIMS), in the form of a custom electrospray laboratory diagnostic and an analytical-grade system, yields a comprehensive analysis of secondary ions formed from energetic ion beam impacts with ionic liquid thin-film substrates. Results revealed nearly identical positive secondary ion species for both EMI-BF$_4$ and EMI-Im thin films, whereas EMI-Im produced a more diverse set of negative ions consistent with the greater chemical complexity of its anion. The analytical-grade SIMS spectra revealed many relatively high mass-to-charge ratio secondary ions likely below the detection limit for the laboratory diagnostic, thus shifting the average $m/z$ value to above the monomer mass for most spectra. Finally, optical profilometry analysis reveals an estimated 0.5 nm/min sputter rate for an electrospray ion plume bombarding an ionic liquid thin film. 
\end{abstract}

\section{Introduction}
Electrospray thrusters form a distinct class of electric propulsion systems, defined by their use of organic, non-volatile, molten salt propellants known as ionic liquids (ILs) \cite{prince_ionicliquids_2012, lozano_ionic_2005}. Thrust results from the electrostatic emission of ions and charged droplets at the propellant meniscus, generating a polydisperse molecular-ion plume. These micro-electric propulsion devices occupy a niche area within space propulsion, owing to nanonewton-level thrust per emitter \cite{whittaker2025characterization} which is ideal for applications like ultra-precise disturbance rejection for flagship-class missions like the Laser Interferometer Space Antenna (LISA) \cite{PhysRevD.98.102005} and the Habitable Worlds Observatory \cite{gaudi2020habitableexoplanetobservatoryhabex}. In addition, their inherently low mass, volume, and power requirements, coupled with mechanical simplicity, render electrospray thrusters particularly well suited as primary propulsion for small satellites \cite{lemmer2017propulsion}.

However, demonstrated operational lifetimes of ionic liquid electrospray systems (500 to 3600 hours) are orders of magnitude below many mission requirements \cite{uchizono_role_2021, Collins2019GridImpingement}. Plume–surface interactions are believed to underlie the primary lifetime-limiting mechanisms across electrospray operating modes (pure-ion and droplet), occurring via two primary pathways: intrinsic processes such as plume overspray\cite{thuppul2020lifetime}, and facility-induced effects arising during ground testing\cite{uchizono_facility_2022}. There is a wide gap in the fundamental knowledge of how the relatively complex molecular nature of the ionic-liquid electrospray plumes interact with exposed surfaces compared to traditional noble-gas electric propulsion propellants, and how these fundamental surface-interaction processes contribute to premature failure pathways.

\begin{wrapfigure}{r}{0.55\textwidth}  
    \centering
    \includegraphics[width=0.99\linewidth]{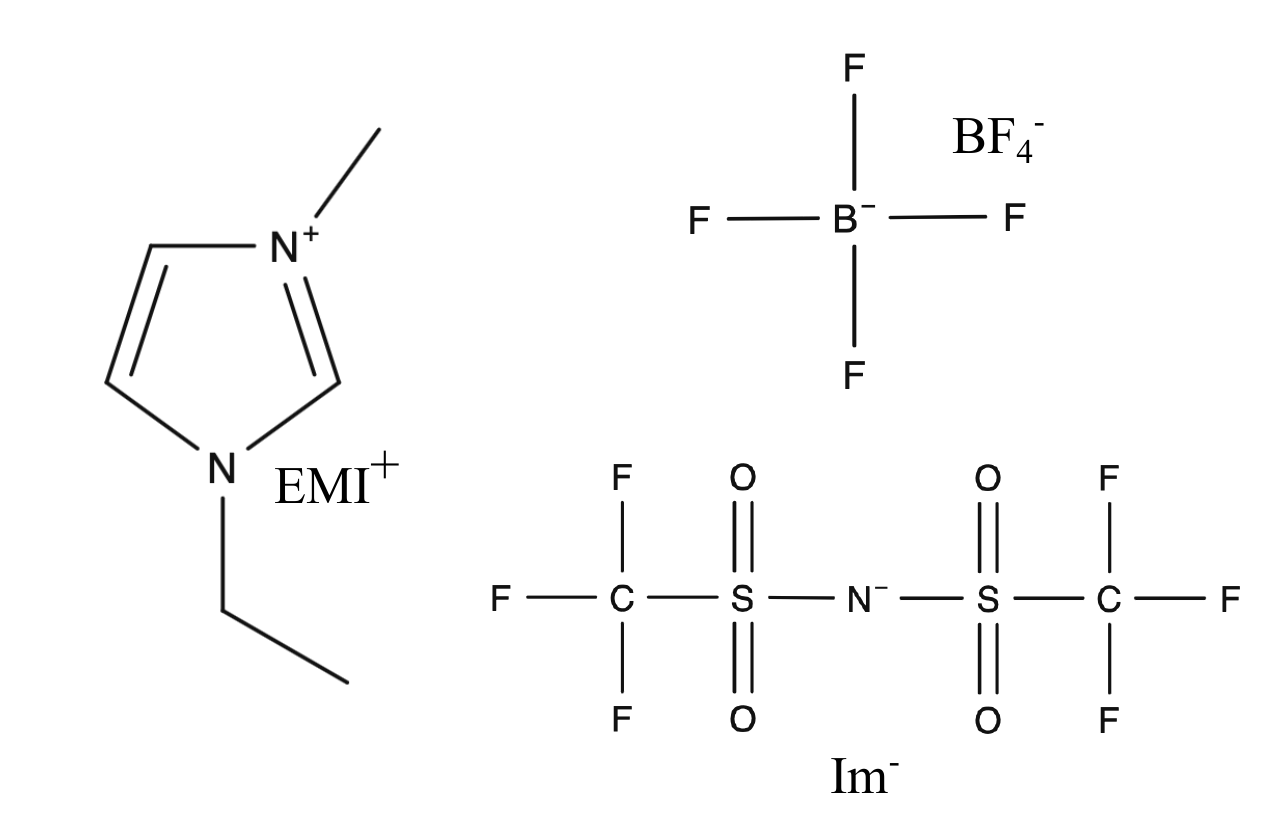}
    \caption{Structural diagrams of the cation EMI$^+$ ($m/z$ = 111 amu), as well as anions BF$_4^-$ ($m/z$ = 87 amu) and Im$^-$ ($m/z$ = 280 amu).}
    \label{fig:ILs}
\end{wrapfigure}

Ionic liquid electrospray plumes are inherently polydisperse -- comprised of distinct ions, solvated neutrals, oligomeric clusters and droplets spanning a wide range of velocity, mass-to-charge ratio, and energy distributions, with common propellant constituent ions shown in Fig. \ref{fig:ILs}. In droplet-mode operation—favored for enhanced thrust-to-power—cone-jet emission produces charged droplets spanning a range of charge-to-mass ratios, whereas pure-ion-regime (PIR) sources deliver plumes dominated by discrete ions and solvated neutrals, yielding superior specific impulse. In practice, both regimes exhibit some degree of mixed emission; droplet-mode plumes contain discrete ions, while PIR operation often includes a portion of droplets. This diversity of constituent states convolutes the investigation of plume-surface interactions. When an energetic electrospray plume comprised of both ions and droplets impacts an exposed surface, processes spanning from secondary electron and ion emission to desorption and splashing occur dependent on distinct impacting species and local target morphology \cite{uchizono2021role}. The confluence of these processes form what is known as secondary species emission (SSE), which encompasses any ejecta formed from plume-surface collisions in the form of neutrals, ions, clusters, or droplets \cite{uchizono_facility_2022, hofheins2025electrospray}. These secondary species interact with the existing system and influence diagnostic measurements, facility effects and lifetime qualification of electrospray thrusters \cite{kerber2025characterization, lyne2024quantifying}. 

\begin{figure}
    \centering
    \includegraphics[width=1\textwidth]{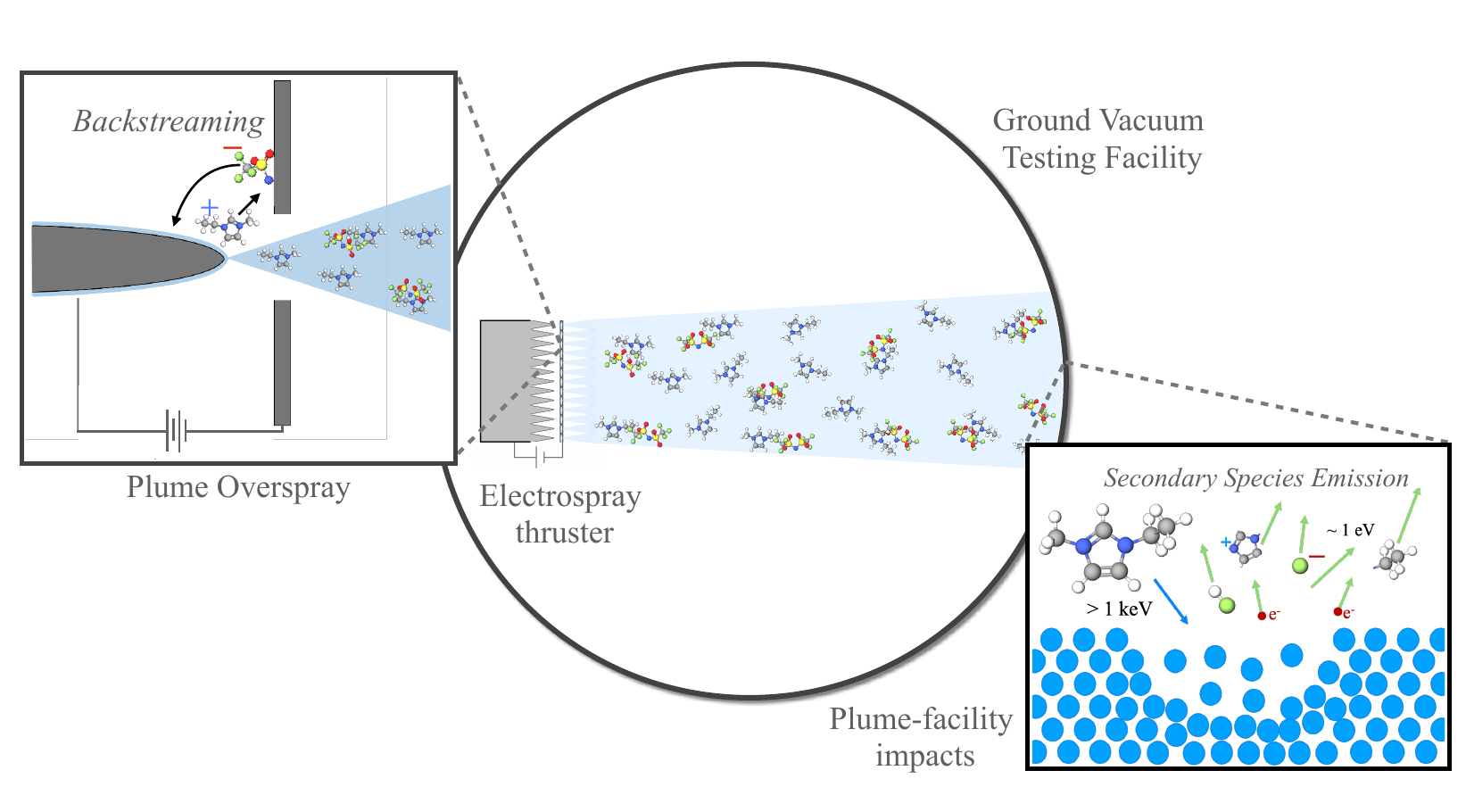}
    \caption{Electrospray operation involves molecular ion plumes impacting extractor electrodes and downstream facility surfaces, causing the formation of secondary species. These secondary species can backstream to the emitter and introduce great measurement uncertainty.}
    \label{fig: sseprocesses}
\end{figure}

In electrospray systems, a propellant-wetted emitter is aligned concentrically with a downstream extractor grid to establish an electric field sufficient to overcome propellant surface tension, thus enabling ion emission from the liquid interface. Even in ideal emitter/extractor geometries, intrinsic emission properties lead to plume impingement of the extractor electrode as shown in Fig. \ref{fig: sseprocesses}. In the PIR, neutral particles formed from both dissociation in acceleration region\cite{miller2020measurement, petro2022multiscale} and collisions between particles in the plume\cite{Smith2024collisions} have large velocity and angular distributions as they are not guided by electric field. Overspray causes propellant accumulation and electrochemical reactions on the extractor at low impact energies \cite{uchizono_role_2021}. At higher energies, it drives sputtering-induced electrode degradation and secondary electron or ion emission \cite{hofheins2025electrospray, krejci2017emission, uchizono_role_2021}. Droplet mode systems identify overspray directly as most significant failure mechanism as propellant accumulates on downstream electrodes, leading to electrical shorts and backspray of opposing firing polarity particles \cite{thuppul2020lifetime, Collins2019GridImpingement}. However, practical implementations of electrospray thrusters include arraying hundreds\cite{petro2020characterization, villegas2024emission,natisin2020fabrication} if not thousands\cite{whittaker2025characterization} of individual emitters into a singular thruster “chip” for increased thrust throughput. At this micron level scale, machining and alignment tolerances can exacerbate propellant-extractor impacts and aforementioned lifetime limiting processes \cite{whittaker2024statistical}.

“Facility effects” refer to interactions between a propulsion test article and terrestrial testing facilities. Accounting for and mitigating these influences is essential to isolate thruster-intrinsic performance from chamber-induced processes such as electrical coupling and plume–wall interactions. While facility effects in traditional noble-gas plasma propulsion have been extensively characterized over decades \cite{lobbia2019accelerating, tran2024carbon, rosenberg1962sputtering}, analogous effects in ionic-liquid electrospray systems remain comparatively underexplored with studies such as those by Uchizono, Klosterman, Kerber, Lyne, and Shaik et al. \cite{uchizono_dissertation_nodate, klosterman_ion-induced_2022, kerber2025characterization, lyne2024quantifying, shaik2024RGA} contributing to this emerging body of work. Uchizono et al. \cite{uchizono2021role} reported that plume impingement on vacuum-facility surfaces generates significant secondary species emission (SSE), with Geiger et al. \cite{geiger2025secondary} showing that even low-energy plume impacts produce measurable SSE. The resulting backstreaming particles can drift toward components of opposite polarity, interacting with both the electrospray emitters in the acceleration region and the extractor electrode in the field-free region. These backstreaming currents introduce significant uncertainty in performance and lifetime measurements by altering the physical properties of the ionic liquid and contributing to anomalous mass loss through SSE-induced Ohmic dissipation \cite{uchizono_role_2021}. Extractor degradation commonly exhibits decomposed propellant on the facility-facing surface \cite{krejci2017emission}, with a hypothesis that this accumulation originates from backstreaming species generated by plume–wall impacts. Significant work at NASA Jet Propulsion Laboratory in the form of optimized beam-target design for electrospray facility testing underscores the importance of mitigating backstreaming particles for lifetime qualification and to mitigate uncertainty in diagnostic measurements \cite{arestie2025ionic}, but even optimal beam targets with appropriate biasing still accelerate any orthogonal positive secondary species directly back to the electrospray system \cite{uchizono_facility_2022}.

Previous efforts on investigating plume-surface interactions specific to electrospray plumes include quantifying secondary species yields \cite{klosterman_ion-induced_2022, uchizono_positive_2022}, and energy distributions \cite{uchizono_positive_2022}, as well as probing luminescence effects from facility and extractor impingement\cite{kerber2025characterizationpt1, kerber2025characterization}. The luminescence work by Kerber et al.\cite{kerber2025characterizationpt1, kerber2025characterization}, in addition to work by Shaik et al., \cite{shaik2024RGA} who quantified neutral by-products from electrospray plume impingement of stainless steel via residual gas analyzer, investigated neutral species formed from electrospray-plume impacts. These results have been qualified by molecular dynamics studies of discrete IL ion impacts with potential and gold walls \cite{ximo_modellingsurface, bendimerad2022molecular}. However, the first experimental study into the chemical composition of charged secondary ions formed from plume-surface interactions were enabled by the design of the novel laboratory diagnostic, electrospray time-of-flight secondary ion mass spectrometry (ESI TOF-SIMS) \cite{hofheins2025electrospray}. The quantification of the mass-to-charge ratio of charged secondary species is particularly important for quantifying exactly what species, in distinct relative intensities, are being accelerated back to the emitter through present electric fields and extractor electrode during electrospray testing and operation and likely heavily responsible for lifetime limiting degradation. These initial tests with ESI TOF-SIMS diagnostic showed that a quasi-pure ion regime plume with a subset of droplet population ($\sim$10-15\% by current fraction) impingement of a pure silver target forms a wide range of atomic, molecular, and a small subset of continuous m/z secondary ions of both polarities related to ionic liquid ion fragments, the target material, and surface contamination \cite{hofheins2025electrospray}.

As stated, the goal is for electrospray thrusters to demonstrate stable thrust output for thousands to tens of thousands of hours. In vacuum facility ground-testing, 100\% of the beam will be intercepted by a downstream surface. In droplet-mode thrusters, the lower specific energy of droplets leads to propellant accumulation on extractor electrodes and facility beam targets \cite{ziemer2007flight, arestie2025ionic}. In PIR systems, plume species including low energy neutrals and typically a subset of low mass-to-charge ratio droplets may contribute to propellant deposition on similar surfaces \cite{petro2022multiscale, krejci2016scalable}. Therefore, it is important to characterize not only secondary species that result from plume impacts with bare surfaces, but with propellant thin films as well. Initial study of quasi-pure ion regime plumes presented many atomic and molecular secondary ions likely related to ionic liquid ion fragments and are likely present from re-sputtered deposited propellant by the plume during the tests \cite{hofheins2025electrospray}. In addition, luminescence work showed many ionic-liquid related secondary species formed either from collision-induced dissociation at the impact site or from deposited propellant liberated upon impact \cite{kerber2025characterizationpt1, kerber2025characterization}. Finally, utilizing a thermally controlled quartz crystal microbalance to measure backstreaming particles from a beam target, Arestie et al. \cite{arestie2025ionic} suggested that the higher measured mass-flux accumulation may result from plume impingement on deposited propellant rather than bare surfaces. Secondary species related to plume impacts with deposited propellant is prolific among all electrospray plume operating modes, and is important for both overspray and facility effect considerations, yet the specific ejecta from these collisions have not been studied. 

Time-of-flight secondary ion mass spectrometry (TOF-SIMS) is a widely adopted surface-analysis technique. It employs a primary ion beam that ranges from monoatomic ions to gas-cluster ions containing thousands of atoms to bombard a target of interest \cite{van_der_heide_secondary_2014}. These collisions induce sputtering, the removal of material as neutrals and as secondary ions of both polarities \cite{van_der_heide_secondary_2014}. A mass-spectral representation of the extracted secondary ions provides insight of the sample chemical composition. TOF-SIMS studies using atomic and small metal-cluster primary beams to study ionic-liquid thin films, notably EMI-Im, have been reported \cite{GUNSTER20083403, BUNDALESKI201319}. Fujiwara et al. \cite{fujiwara_time--flight_2014, fujiwara2014development} demonstrated that a droplet-mode ionic liquid electrospray can function as a TOF-SIMS primary ion source demonstrated by analysis of an analogous IL thin film. However, there is a gap in the literature of resultant ejecta from complex, quasi-pure ion regime electrospray plume impacts with ionic liquid thin films. 
 
This work leverages both a recently developed electrospray ESI TOF-SIMS laboratory diagnostic \cite{hofheins2025electrospray} and a commercial TOF-SIMS system to investigate secondary ions generated by bombardment of ionic liquid propellant thin films. Thin films of two widely used electrospray propellants—1-ethyl-3-methylimidazolium tetrafluoroborate (EMI-BF$_4$) and 1-ethyl-3-methylimidazolium bis(trifluoromethylsulfonyl)imide (EMI-Im), shown in Fig.\ref{fig:ILs} -- were exposed to both high-energy molecular ion electrospray plumes and small metal cluster ion plumes. The resulting spectra enable analysis of chemical composition, relative abundances, and sputter rates arising from plume–propellant interactions, and provide a direct comparison with high–mass-resolution commercial TOF-SIMS of ionic liquid thin films.

\section{Methods}
\subsection{ESI-TOF SIMS}
\begin{figure}
    \centering
    \includegraphics[width=.75\textwidth]{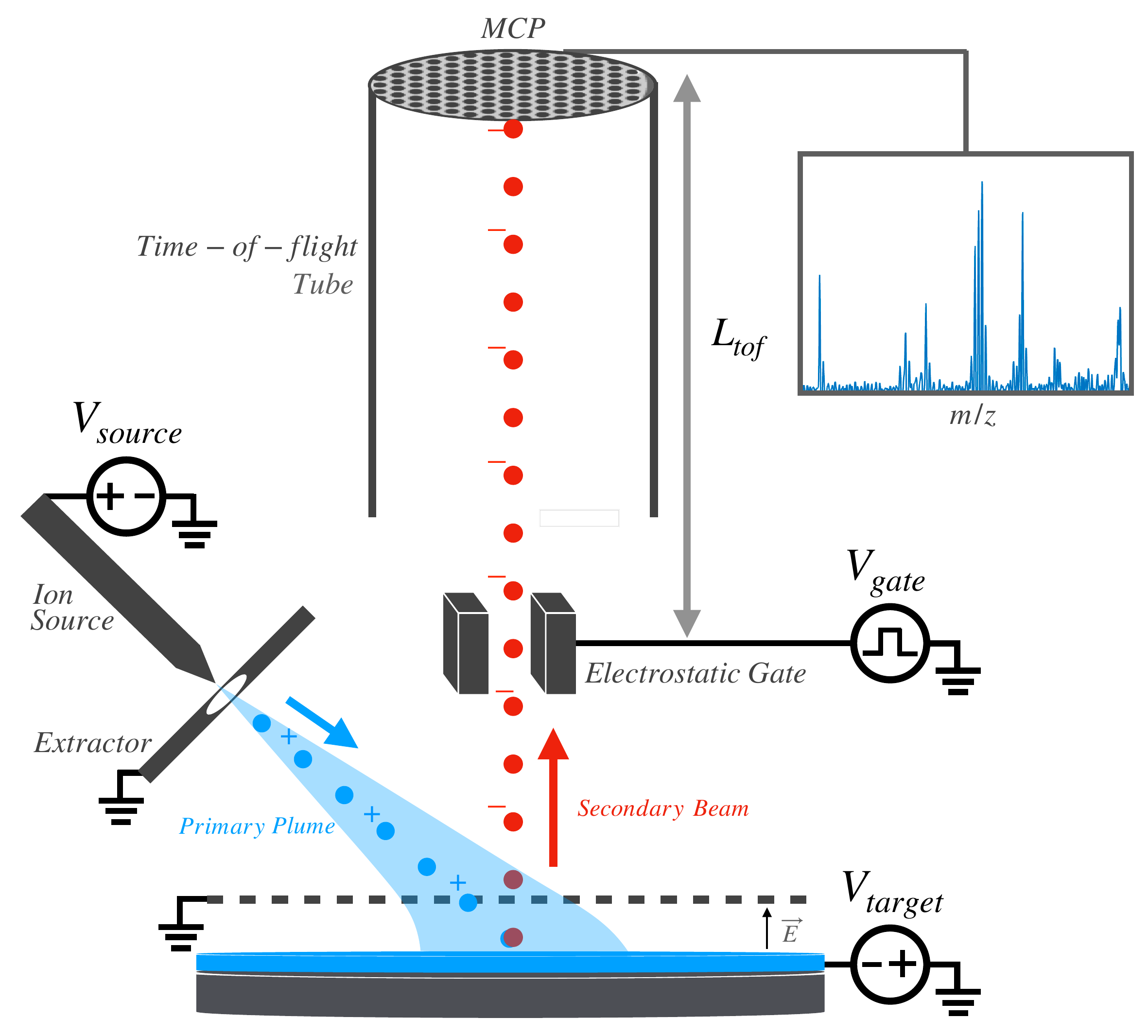}
    \caption{Electrospray TOF-SIMS diagnostic, where a single electrospray ion source directs a polydisperse molecular ion plume at a target of interest to induce sputtering and the formation of secondary ions to then be analyzed via a linear time-of-flight mass spectrometer.}
    \label{fig: simsdiagram}
\end{figure}

An electrospray propulsion-specific TOF-SIMS laboratory diagnostic as described in Hofheins, et al. \cite{hofheins2025electrospray}, and shown in Fig. \ref{fig: simsdiagram}, which allows for dual-polarity analysis of secondary ions formed from plume-surface impacts. The basis of the diagnostic is to fire an externally-wetted tungsten single emitter electrospray ion source at a target of interest in order to induce sputtering and selectively extract the secondary ions from the sputtered `cloud'. These secondary ions are thus analyzed via a linear time-of-flight mass spectrometer to produce a mass spectrum representative of the secondary species population of a given polarity. 

\subsubsection{Ion Source}

The electrospray ion source employed consisted of a single, externally-wetted tungsten needle with full details in Ref. \cite{hofheins2025electrospray}. Briefly, a tungsten rod was electrochemically etched via the process described in Lozano et al. \cite{lozano_ionic_2005} to a 8.6 $\mu$m radius of curvature as determined by scanning electron microscope imaging analysis. This ion source was mounted in polyether ether ketone (PEEK) and polytetrafluoroethylene (PTFE) housing and mounted to a goniometer for in-situ firing angle control \cite{ulibarri2025direct}. Room tempterature ionic liquid propellants 1-ethyl-3-methylimidazolium tetrafluoroborate (EMI-BF$_4$) and 1-ethyl-3-methylimidazolium bis(trifluoromethylsulfonyl)imide (EMI-Im) are loaded into the reservoir by pipetting 4-5 $\mu$L of propellant over the needle tip. A stainless steel extractor plate with a 2 mm diameter aperature aligns concentrically with the ion source needle tip. By applying a high voltage to the propellant reservoir and grounding the extractor, a polydisperse ion plume forms with a maximum energy equivalent to the applied voltage, $V_{source}$. 

In the quasi pure-ion regime, the plume is composed primarily of cations in positive mode and anions in negative mode, along with solvated neutrals (Table \ref{tab: solvationorder}). Species take the general form [C$^+$]$_n$[C$^+$A$^-$] in positive polarity and [A$^-$]$_n$[C$^+$A$^-$] in negative polarity. Here, $n$ = 0 corresponds to a monomer, $n$ = 1 a dimer, and $n$ = 2 a trimer, with higher $n$ representing larger ion–neutral clusters of a given polarity. However, emission from these and many other `PIR' sources typically possess larger mass-to-charge ratio nanodroplet constituents ranging out to $m/z$ = 10$^5$, representing as much as 20\% of the total current fraction.

The characteristic polydispersity of electrospray plumes leads to fragmentation in both the acceleration and field-free regions, producing broad plume energy distributions. These distributions span from the maximum energy of $V_{source}$ down to much lower energies associated with room-temperature neutrals and dissociated solvated ions \cite{Lozano_2006, ma2021plume, miller2020measurement}. Typical beam divergences are on the order of 20$^\circ$ \cite{petro2022multiscale}. 

\begin{table}[!ht]
    \centering
    \caption{Plume species and solvation order.}
    \label{tab: solvationorder}
    \begin{tabular}{lllll}
    \hline\hline
        \textbf{Plume Species} & \textbf{Polarity} & \textbf{Solvation Order, n}  &\textbf{Mass [amu]} \\ \hline
        EMI$^+$(EMI-BF$_4$)$_n$ & + & 0 & 111 \\ 
        ~ & ~ & 1   & 309 \\ 
        ~ & ~ & 2  & 507 \\ 
        BF$_4^-$(EMI-BF$_4$)$_n$ & $-$  & 0  & 87 \\ 
        ~ & ~ & 1 & 284 \\ 
        ~ & ~ & 2   & 482 \\ 
        EMI$^+$(EMI-Im)$_n$ & $+$ & 0  & 111 \\ 
        ~ & ~ & 1  & 502 \\ 
        ~ & ~ & 2  & 894 \\ 
        Im$^-$(EMI-Im)$_n$ & $-$  & 0  & 280 \\ 
        ~ & ~ & 1  & 671 \\ 
        ~ & ~ & 2   & 1063 \\ \hline\hline 
    \end{tabular}
\end{table}

\subsubsection{Ionic Liquid Thin Film Target}
The target apparatus consists of a 4.5" diameter aluminum disk mounted to a standoff by a PEEK plate to maintain electrical isolation. This disk is equipped with copper SEM clips to mount 4" silicon wafer targets. A concentric aluminum ring with a spot-welded high-transparency stainless-steel mesh (88\%, TWP inc) mounts in front of the target via 0.25" PEEK spacers to act as a secondary ion acceleration grid. By applying a high voltage, $V_{target}$, to the target and grounding the acceleration grid, an electric field forms to both suppress secondary ions of one polarity and accelerate the other orthogonal to the surface. The maximum energy of the collision is the combined energy gained from the source and target acceleration regions, and measures 

\begin{equation}
    E_{impact, max} = |\pm V_{source}| +|\mp V_{target}|. 
\end{equation}

\noindent This is denoted as the maximum impact energy due to the presence of slower species from fragmentation and neutral constituents of the plume. The nominal magnitudes of the source and target voltages is 1.5 kV and 2.5 kV respectively, producing a maximum impact energy of 4 keV. However, the ion source emission characteristics vary between propellants and emission polarities. Namely, the source operating with EMI-BF$_4$ in the negative polarity exhibits more stable firing behavior at $V_{source}$ = $-$2 kV. To compensate, the target was held at $+$ 2 kV to maintain equivalent 4 keV impact energy across tests.

\begin{figure}
    \centering
    \includegraphics[width=1\textwidth]{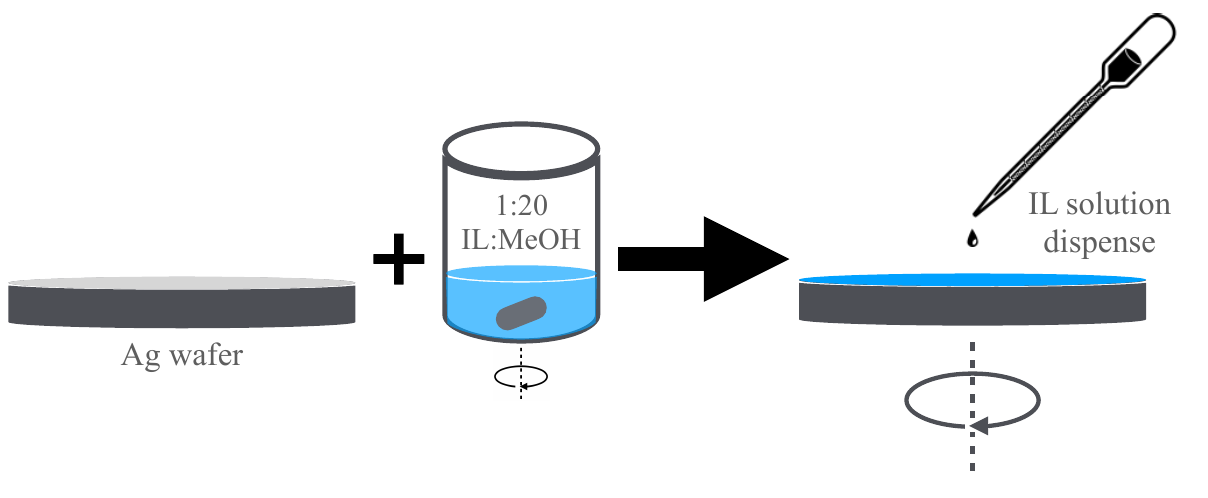}
    \caption{Ionic liquid thin film spincoat process, where 3 mL of an ionic liquid/methanol mixture was dispensed on a silver-coated silicon wafer.}
    \label{fig: spincoat}
    
\end{figure}

The targets for these tests are silver-coated 4" silicon wafers spin-coated with ionic liquid propellant, with the process depicted in Fig. \ref{fig: spincoat}. First, silicon wafers were coated with 100 nm of pure silver via electron beam evaporation (CVC4500) at the Cornell NanoScale Facility. To make the propellant thin films, a solution of 1:20 mass ratio of ionic liquid to methanol solution was prepared and continouslly mixed over a 24 hour period to ensure homogeneity \cite{boning_ILthinfilm}. Methanol serves as a volatile thinning agent, reducing the viscosity of the ionic liquid to enable the formation of a uniform, continuous wet film. It is expected to fully evaporate during the spin process, leaving behind only the ionic liquid layer. By dispensing $\sim$ 3 mL of the solution onto a slowly spinning silver wafer in a spincoater, and then accelerating to 2000 rpm with 300 ramp speed for 45 s, IL films on the order of 100 nm were created as determined by optical profilometry analysis further discussed in Sec. \ref{sec: opticalprof}. 

\subsubsection{Linear Time-of-flight}
Time-of-flight mass spectrometry (TOF) is a commonly used diagnostic for electrospray propulsion systems to directly identify beam composition and infer performance characteristics \cite{jia-richards_quantificationTOF, larriba2007monoenergetic}. For electrospray TOF-SIMS, the technique is utilized to analyze the mass-to-charge ratio of secondary ions formed from plume-surface impacts. Upon surface collision, a fraction of the sputtered material ejects from the surface as a low energy ($<$ 100 eV) atomic and molecular secondary ions representative of the target material \cite{van_der_heide_secondary_2014}. Subject to the force of the applied electric field, the secondary ions are either suppressed to the surface or accelerated orthogonal to the target. The accelerated secondary ions form an approximately monoenergetic secondary ion beam that is analyzed via the time-of-flight system. 
The linear time-of-flight system employed is described fully in Cogan et al. \cite{cogan_2023_electrospray} and Hofheins et al. \cite{hofheins2025electrospray}, with the main components consisting of a parallel-plate electrostatic deflection gate, a meter long flight tube, and a microchannel plate (MCP) detector. A large potential difference ($+$2 kV) applied and removed from the gate at a given frequency (0.5 -- 1 kHz) allows for alternating deflection and clearance of the ion beam of interest. When the potential difference is removed, this establishes the initial `time' where the monoenergetic particles are allowed to travel down the flight tube. The time that it takes a species of like mass-to-charge ratio particles to arrive at the MCP detector, and thus produce an increase in current, can be related to the mass of the species through: 
\begin{equation}
    t_{TOF} = L_{TOF}\sqrt{\frac{m}{2qV}} + t_{delay}
\end{equation}

\noindent where $t_{TOF}$ is the flight time of the given species, $t_{delay}$ recognizes any time delay introduced by the system, $L_{TOF}$ is the flight path measured from the center of the electrostatic gate to the face of the MCP, $m$ is the mass of the species, $q$ is the elementary charge and $V$ is the acceleration voltage, either $V_{source}$ for TOF on the primary plume or $V_{target}$ for TOF-SIMS analysis. The collected current is further multiplied by a transimpedance amplifier (0.5 $\mu$A/V) and recorded and averaged on an oscilloscope. The result is a time-of-flight curve, where each step in current represents a higher mass-to-charge species reaching the detector. The derivative of this curve gives a typical mass spectral representation of the detected species, with peak height corresponding to relative intensity. The flight distance for primary TOF measured $L_{TOF}$ = 1.01 m while the TOF-SIMS distance measured 0.957 m with an average time delay of 155 ns, with further curve fitting analysis described in previous work \cite{hofheins2025electrospray}. 

\subsection{IONTOF TOF-SIMS5}
To compare and identify secondary species produced by ionic liquid electrospray plume impacts, data was collected using an analytical-grade TOF-SIMS system equipped with a liquid metal ion gun (LMIG) primary beam. Specifically, a TOF-SIMS5 (IONTOF GmbH) at the New York Center for Research, Economic Advancement, Technology, Engineering and Science (NY CREATES) was used to analyze secondary ions generated by Bi$_3^+$ cluster impacts on thin films of EMI-BF$_4$ and EMI-Im.

This commercial system utilizes a pulsed Bi$_3^+$ primary ion beam and a reflectron time-of-flight analyzer, achieving mass resolving power up to 6000 ($m/\Delta m$); for example, at $m = 111$ amu, the system can resolve peaks separated by just 0.02 amu. Thin film samples were prepared by spin-coating silver-coated 1 cm $\times$ 1 cm silicon wafer coupons with identical ionic liquid formulations and procedures used for the larger 4" wafers.

Canonically, TOF-SIMS analysis is divided into two main data collection modes: static SIMS, which limits primary ion dose to ensure the detected secondary ions originate from an undisturbed surface region, and dynamic SIMS, which involves sputtering through multiple layers to enable depth profiling \cite{van_der_heide_secondary_2014}. All data in this study were collected in static mode using a 25 keV Bi$_3^+$ beam with a 5 $\mu$m spot size and 0.11 pA current, rastered over a 150 $\mu$m$^2$ region of interest.

Small-cluster primary ion beams like Bi$_3^+$ are known to induce softer ionization compared to atomic beams, due to their lower energy per individual ion and reduced energy density at the impact site. This often leads to the detection of larger, intact molecular fragments, rather than small, highly fragmented species typically observed with small atomic ion sources \cite{dubey_comparisonBi, NAGY2007144}. As a result, the Bi$_3^+$ beam provides a closer analogue to the polydisperse molecular ions characteristic of electrospray plumes for comparison purposes.

\section{Results}
\label{sec: results}
\subsection{Primary Plume Composition}

\begin{figure}[h!]
    \centering
    \includegraphics[width=1\textwidth]{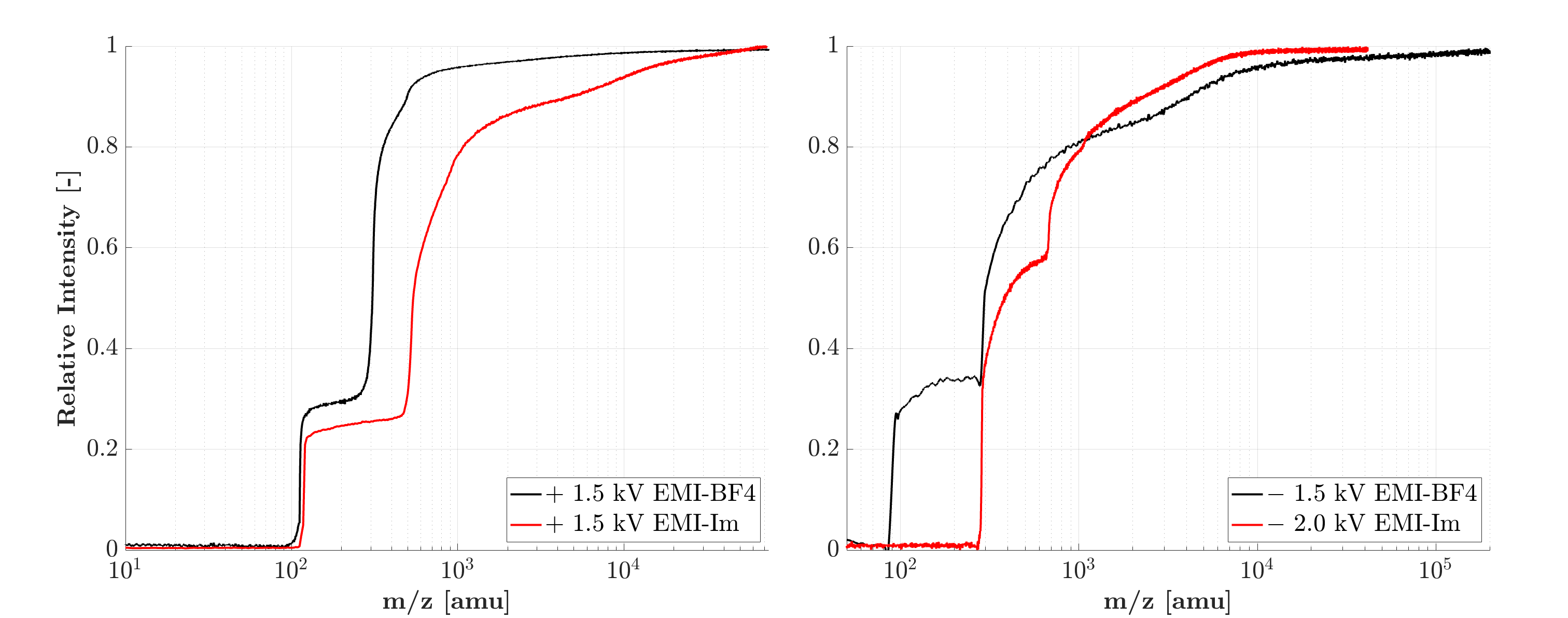}
    \caption{Primary TOF showing a plume composed of primarily ions, with a portion of continuous small droplets for both EMI-Im and EMI-BF$_4$ propellants.}
    \label{fig: primarytof}
\end{figure}

The dual-polarity plumes generated by the primary electrospray source were characterized via TOF for both IL propellants, as shown in Fig.~\ref{fig: primarytof}. To ensure consistent impact energy across emission modes in the subsequent SIMS experiments, TOF measurements were acquired at equal-magnitude firing voltages ($V_{\text{source}} = \pm 2~\text{kV}$). An exception was required for the negative EMI-BF$_4$ mode, where unstable emission was observed at $-2~\text{kV}$ and therefore measurements were taken at $-1.5~\text{kV}$. To maintain comparable plume impact energies under these conditions, the target was biased to a proportionally higher potential during ESI TOF-SIMS tests involving the negative EMI-BF$_4$ plume. Each TOF trace represents 65{,}000 waveform averages on the oscilloscope, with the negative EMI-BF$_4$ curve additionally processed using a 20-point Savitzky--Golay filter~\cite{savitzky1964smoothing} applied to data sampled at 14.5~MHz to mitigate reduced signal-to-noise.

The primary electrospray plumes all reflect typical ionic liquid compositions produced by externally wetted and porous emitters operating near the pure-ion regime (PIR) \cite{lozano_ionic_2005}. In this regime, the plume consists predominantly of ion species in the form of monomers and dimers (n = 0,1), with a smaller fraction (less than 20\% by current fraction) composed of higher mass-to-charge ($m/z$) solvated neutrals and nanodroplets. The positive polarity plumes from EMI-BF$_4$ and EMI-Im exhibited comparable monomer fractions (approximately 25–30\%). However, EMI-Im showed proportionally fewer dimers and trimers, accompanied by a larger continuous $m/z$ droplet tail (15\%) compared to EMI-BF$_4$ (8\%).

Negative polarity operation similarly produced characteristic plume distributions for these ion sources. The EMI-BF$_4$ plume consisted of roughly equal parts monomers, dimers, and higher $m/z$ species. In contrast, the negative EMI-Im plume was composed primarily ($\sim$85\%) of solvated ions, with the remaining 15\% attributed to a broader droplet tail.

Overall, the primary electrospray source operated in a typical quasi-pure ion regime for both polarities and propellants, with the majority of emitted species appearing as single cations or anions and solvated ion clusters, along with a distinct fraction of higher $m/z$ droplets.

\subsection{Positive Secondary Species}
The positive secondary-ion mass spectra produced from negative-plume impacts at a maximum impact energy of 4~keV on IL thin films are shown in Fig.~\ref{fig:combinedPosSIMS}, with the detected peaks and their relative abundances listed in Table~\ref{tab:positive species}. For direct comparison, the corresponding mass spectra obtained from 25~keV Bi$_3^+$ bombardment of the IL films using an analytical-grade IONTOF TOF-SIMS5 are also shown in Fig.~\ref{fig:combinedPosSIMS}, mirrored across the $y$-axis. All spectra are independently normalized to their respective most intense peak..

\subsubsection{ESI TOF-SIMS Spectra}
\begin{figure}[h!]
    \centering
    \begin{subfigure}[t]{0.78\textwidth}
        \centering
        \includegraphics[width=\textwidth]{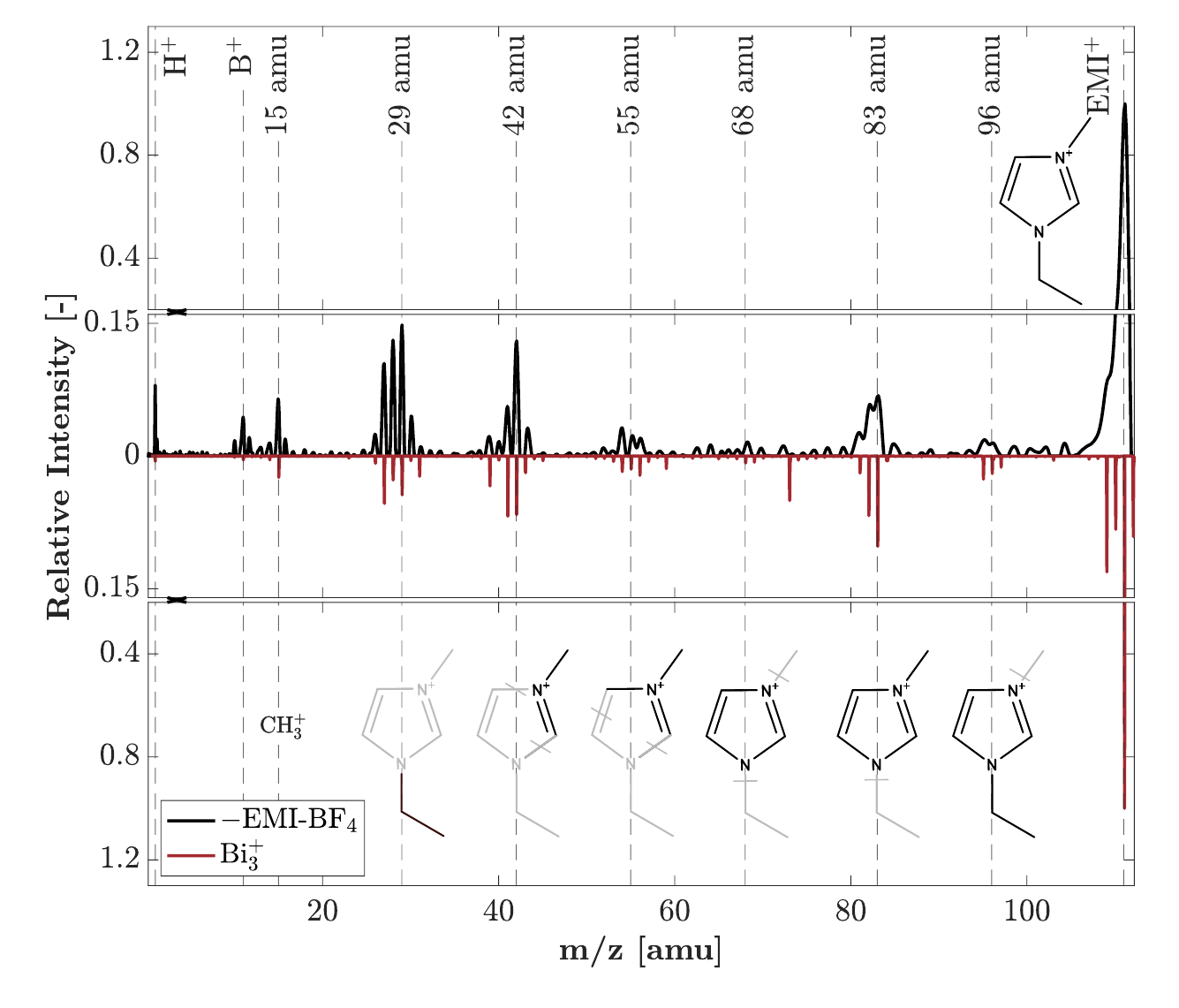}
        \caption{EMI-BF$_4$ thin film.}
        \label{fig:posEMIBF4}
    \end{subfigure}
    
    \vspace{0.75em} 
    
    \begin{subfigure}[t]{0.78\textwidth}
        \centering
        \includegraphics[width=\textwidth]{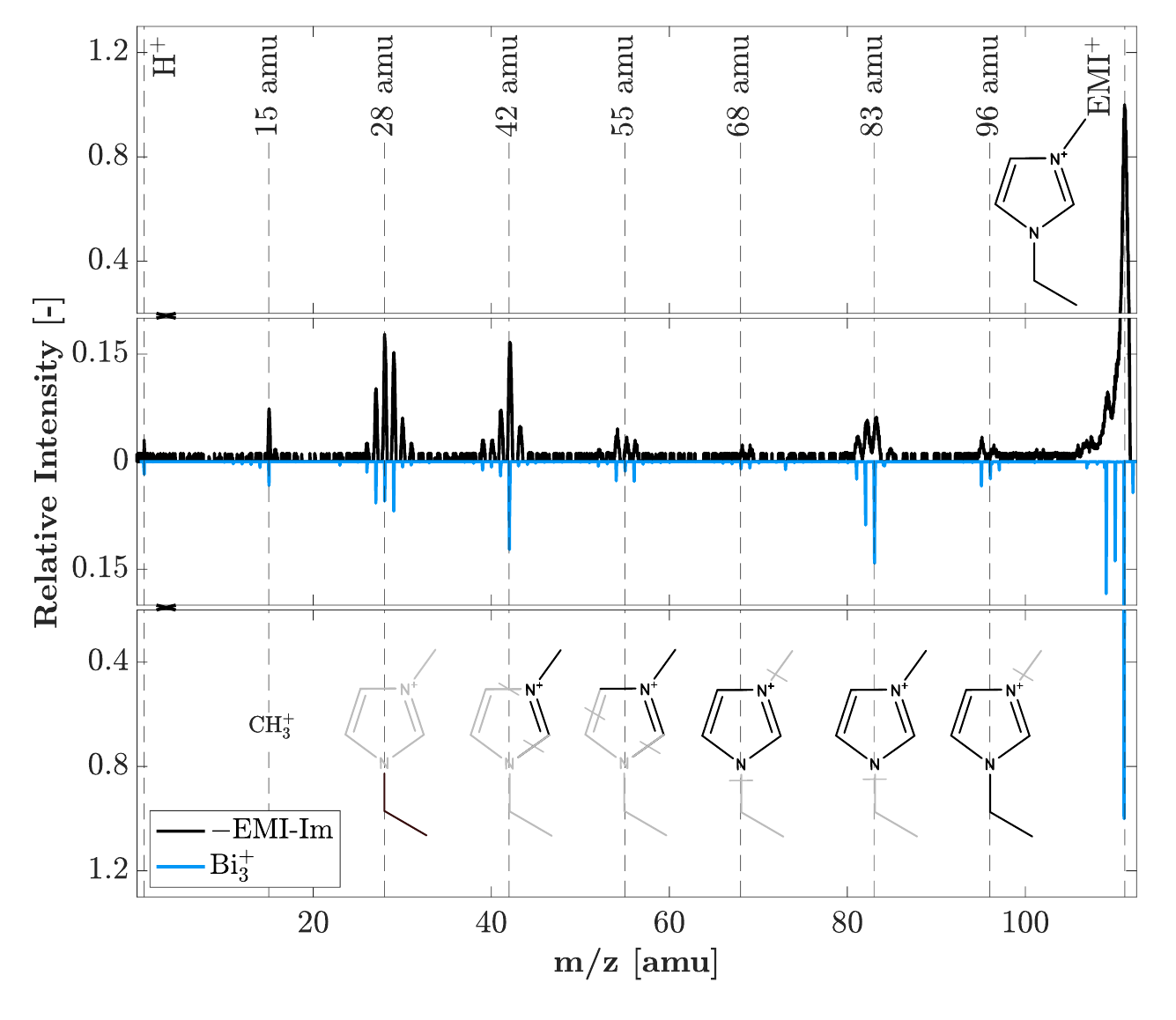}
        \caption{EMI-Im thin film.}
        \label{fig:posEMIIM}
    \end{subfigure}
    
    \caption{Positive secondary ion mass spectra of EMI-BF$_4$ and EMI-Im thin films exposed to (top of each subfigure) a 4 keV $E_\text{impact}$ negative electrospray plume, and (bottom) a 25 keV Bi$_3^+$ primary ion beam in a commercial TOF-SIMS system. All spectra are normalized to the most intense peak. Structural diagrams of the EMI$^+$ cation fragments are shown at the corresponding mass in each spectrum.}
    \label{fig:combinedPosSIMS}
\end{figure}

The positive secondary ion spectra from the electrospray TOF-SIMS diagnostic for both EMI-BF$_4$ (Fig. \ref{fig:combinedPosSIMS}a) and EMI-Im (Fig. \ref{fig:combinedPosSIMS}b) thin films are nearly identical in terms of both detected species and relative intensities due to identical cations. While the figures only show the spectra to $m/z$ 112 amu, no distinct larger species are observed with the exception of the EMI-BF$_4$ and EMI-Im dimers, most clearly seen in the raw TOF curves in the appendix (Fig.\ref{fig: appendixquadplot}).

\begin{wrapfigure}[22]{r}{0.55\textwidth}
    \vspace{-10pt}
    \centering
    \includegraphics[width=0.99\linewidth]{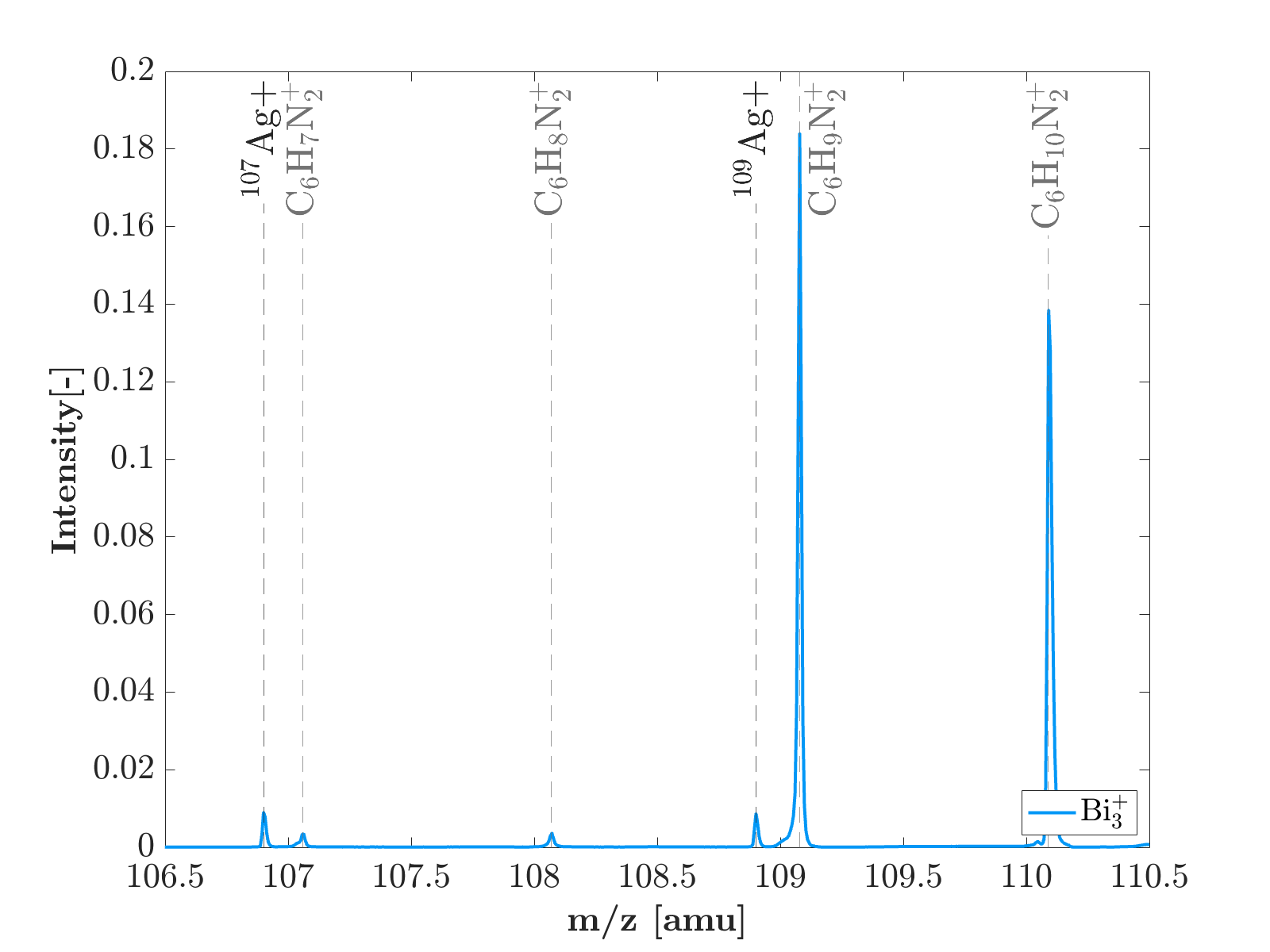}
    \caption{A zoomed portion of the positive TOF-SIMS5 spectrum shown fully in Fig. \ref{fig:combinedPosSIMS}, distinguishing the small silver detections (106.9 and 108.9) from dehydrogenated EMI$^+$ (C$_6$H$_{11}$N$_2^+$).}
    \label{fig: posEMIIMalbanyzoom}
\end{wrapfigure}

The most intense positive secondary ion for both EMI-BF$_4$ and EMI-Im is the intact EMI$^+$ cation at $m/z$ = 111.1 amu, accounting for 44\% and 51\% of the total collected signal. However, spectral structure from $m/z$ = 107 to the EMI peak is also observed. On initial inspection, the peaks at 107 and 109 first point to silver ion detections, as clearly observed as the most intense peaks when an electrospray plume bombards a bare silver target \cite{hofheins2025electrospray}. An alternative hypothesis is that these peaks are EMI$^+$ satellite peaks formed from successive hydrogen losses. Three observations support the latter more than the former. The first is that the small, broad peak observed in the electrospray TOF-SIMS spectra for both IL's at 107 is clearly less intense than the peak at 109 amu, which is inconsistent with the nearly equal isotopic abundance of $^{107}$Ag and $^{109}$Ag. The second is the fact there is an additional inflection point in the dominant EMI$^+$ peak occurring at 110 amu, which is inconsistent with any silver-derived species. Finally, the high mass resolution data from the TOF-SIMS5 system can be leveraged to distinguish between silver ions and dehydrogenated EMI$^+$ ions. The subset of the positive EMI-Im SIMS spectrum between 106 and 110 amu is shown in Fig. \ref{fig: posEMIIMalbanyzoom}, and reveals separate peaks that distinguish between silver and EMI$^+$ fragments. There exists equal, small intensity peaks at the exact masses of silver isotopes at 106.9 and 108.9 amu, in addition to peaks at 107.06, 108.07, 109.08, and 110.09 amu representing hydrogen losses (H = 1.01 amu) from the intact EMI$^+$ cation ($m/z$ =111.1 amu). Therefore, the peaks surrounding the intense 111 amu peak in the electrospray TOF-SIMS diagnostic curves likely represent stronger contributions from dehydrogenated EMI$^+$ with some smaller contributions of silver secondary ions. In addition, these small silver peaks may not originate from the bulk film, as the Bi$_3^+$ static SIMS data was acquired by rastering over a 150 µm$^2$ region of interest, where edge effects or voids in the ionic liquid could have exposed the underlying silver surface.

Following the dominant EMI$^+$ peak, there exists subsequent spectral families representing a typical fragmentation fingerprint of the EMI cation as observed in other related studies \cite{Bell2024, BUNDALESKI201319, GUNSTER20083403}. Peaks representing methyl and ethyl group losses from the intact EMI cation exist at $m/z$ = 96, 83 amu respectively, and are unique to SIMS spectra of IL thin films compared to similar experiments with a bare silver target \cite{hofheins2025electrospray}. In addition, small peaks near 68 amu represent the intact imidazolium ring ion with the loss of both functional groups.

Peaks in the $m/z$ = 54-57 amu family as well as the 39-43 both represent two bond scissions of the imidazolium ring. Both spectra show about a 10\% reduction in relative intensity from the 42 amu group to 55 amu, likely due to the fact that cleaving an intact C$_3$NH$_y^+$ fragment from the aromatic ring structure requires breaking a C-N double bond. This is compared to 42 amu group, where there are an increased number of fragmentation pathways to yield C$_2$NH$_y^+$ isomers, many of which do not require breaking a double bond in the imidazolium ring backbone. Strict hydrocarbons likely wouldn't contribute to these peaks, as the longest continuous chain in both the cations and anions is an ethyl group.

Similarly, the second most intense spectral group after EMI$^+$ corresponds to the peaks between 26-29 amu. These are consistent with both ethyl-based hydrocarbons (C$_2$H$_n^+$) and cyanide-containing species (CNH$_m^+$), indicative of C-N bond cleavage of the imidazolium backbone. The relative higher intensity of these peaks is likely a result of these multiple, similar mass species contributing to single amu peaks. Finally, both spectra have small intensities of atomic hydrogen ions liberated in the sputtering process.

The most distinguishing feature between the positive-ion spectra of EMI-BF$_4$ and EMI-Im appears in the 10--16~amu range. In the EMI-BF$_4$ spectrum, distinct peaks at $m/z = 10$ and 11~amu correspond to boron isotopes generated by anion fragmentation. As expected, these peaks are absent from the EMI-Im spectrum, consistent with the absence of boron in the Im anion. By contrast, the EMI-Im positive spectrum shows no unambiguous Im-derived fragments, since potential positive secondary ions such as CF$_3^+$ at 69~amu cannot be readily distinguished from C$_3$H$_5$N$_2^+$, the ion associated with the imidazolium ring.

\setlength{\intextsep}{1pt}
\vspace{10pt}
\begin{table}
\centering
\caption{\label{tab:positive species} Positive secondary ion collection mode species detection, assuming singly charged.}
\begin{tabular}{cccc}  
\hline \hline
Mass (amu) & Species Identification & \% Abundance EMI-BF$_4$  & \% Abundance EMI-Im \\ \hline
    1 & H$^+$ & 1.8 & 0.6 \\
    10 & $^{10}$B$^{+}$ & 0.8 & --\\
    11 & $^{11}$B$^{+}$ & 2.5 & -- \\
    12 & C$^+$ & 0.9 &  0.2 \\
    13 & CH$^+$ & 0.7 & 0.2 \\
    14 & CH$_2^+$ & 0.7 & 0.3 \\
    15 & CH$_3^+$ & 3.0 & 2.0  \\
    26 & C$_2$H$_2^+$, CN$^+$ &  1.0 &  1.0 \\
    27 & C$_2$H$_3^+$, NCH$^+$ & 4.7 & 4.0 \\
    28 & C$_2$H$_4^+$, NCH$_2^+$& 5.6 &  6.8 \\
    29 & C$_2$H$_5^+$, NCH$_3^+$ & 6.5 & 6.0 \\
    39 & C$_2$HN$^+$, K$^+$ & 1.0 & 0.9 \\
    40 & C$_2$H$_2$N$^+$& 0.5 & 1.0\\
    41 & C$_2$H$_3$N$^+$ & 2.5 & 2.8 \\
    42 & C$_2$H$_4$N$^+$ & 5.8 & 5.3 \\
    52 & C$_3$H$_2$N$^+$ & \textcolor{green}{--} & 0.55 \\
    54 & C$_3$H$_4$N$^+$ & 1.4 & 1.8 \\
    55 & C$_3$H$_5$N$^+$ & 0.9 & 1.1 \\
    56 & C$_3$H$_6$N$^+$& 0.9 & 1.2 \\
    67 & C$_3$H$_3$N$_2^+$ & 0.6 & -- \\
    68 & C$_3$H$_4$N$_2^+$ & 0.3 & 1.0 \\
    69 & C$_3$H$_5$N$_2^+$, CF$_3^+$ & 0.5 & 0.8 \\
    70 & C$_4$H$_8$N$^+$ & \textcolor{green}{--} & \\
    81 & C$_4$H$_5$N$_2^+$ & 1.5 & 2.2 \\
    82 & C$_4$H$_6$N$_2^+$ & 3.0 & 2.9 \\
    83 & C$_4$H$_7$N$_2^+$ & 2.9 & 2.5\\
    94 & C$_5$H$_6$N$_2^+$ & 1.4 & -- \\
    95 & C$_5$H$_7$N$_2^+$ & 0.4 & 2.1 \\
    96 & C$_5$H$_8$N$_2^+$ & 0.8 & 0.6\\
    111 & EMI$^+$ & 44.5 & 51.2 \\
    308 & EMI$^+$(EMI-BF$_4$) &  1.2 & -- \\
    502 & EMI$^+$(EMI-Im) & -- & 0.5 \\
    cont. $m/z$ & -- & 1.6 & 0.4 \\
    \hline \hline
\end{tabular}
\end{table}

\subsubsection{IONTOF TOF-SIMS5 Spectra}
The positive secondary ion spectra from impacts of a 25 keV Bi$_3^+$ beam with EMI-BF$_4$ and EMI-Im thin films on silver substrates are shown to $m/z$ 112 amu in the reflected intensity axis in Fig. \ref{fig: posEMIIMalbanyzoom}. With a mass resolving power exceeding 6,000 (m/$\Delta$m) and a dynamic range spanning more than five orders of magnitude, this system is capable of resolving and quantifying both minor and dominant secondary ion species within a single spectrum. The power of this is revealing secondary ions that the ESI TOF-SIMS diagnostic may fail to detect above the system noise. 

Similar to the positive ESI TOF-SIMS spectra, the positive TOF-SIMS5 spectra show nearly identical species at comparable relative intensities, with the higher mass resolution additionally revealing weaker peaks not observed in the laboratory diagnostic results. In particular, small peaks corresponding to O and F ions appear in both positive spectra, though at very low intensity ($\sim$1/10,000 that of EMI$^+$). While oxygen is present in EMI-Im and fluorine in both anions, these species—as well as H, N, C, and Si—may also reflect background contamination or system memory.
While there is no predictable model for SIMS ion yield, electronegative elements (non-metals like F, O) more easily ionize into negative secondary ions, while electropositive elements (metals like Na, K) tend to ionize more into positive secondary ions. This difference in ionization efficiency is empirically described with a relative sensitivity factor (RSF), which illustrates the wide orders of magnitude variation in ion yields, depending on variables like the ion of interest, the bulk ‘matrix,’ and by polarity \cite{wilson1995sims}. 

In addition, at $m/z$ = 31, the EMI-Im spectra possesses a peak at 30.99 while EMI-BF$_4$ spectra has a distinctly shifted peak at 31.01. This points to a CF$^+$ detection ($m/z$ = 30.99) in the EMI-Im spectra, consistent a fragment of the CF$_3$ structure in the Im anion, while 31.01 in EMI-BF$_4$ more likely points to an HBF$^+$ fragment or CH$_3$OH$^+$ from the methanol solvent. 

Peaks past the dominant EMI$^+$ are present in both spectra. Namely, there are relatively intense peaks around $m/z$ = 125 amu, 137 amu, and 147 amu at 3\% relative intensity and decreasing by a decade per successive group. Both spectra show the respective dimers, with EMI-BF$_4$ trimer appearing as well. Interestingly, while the EMI-Im trimer does not appear directly at $m/z$ = 893 amu, there are even higher mass spectral families that exist out to 970 amu with a peak intensities 5 orders of magnitude below the EMI$^+$ dominant peak. Interestingly, there is a clear spectral pattern below the EMI-BF$_4$ dimer of successive spectral families spaced by 15 amu apart, representing secondary ions formed from methyl losses from the dimer. 

The average secondary ion mass for each spectra was determined from the spectral centroid, calculated as the mass-weighted mean of the intensity distribution across all $m/z$ values. Only peaks 3 times above the mean baseline noise level were accounted for in the integration. With this, the average secondary ion mass was 123 amu for EMI–BF$_4$ and 138 amu for EMI–Im. The higher average mass for EMI–Im reflects the larger contribution of high-$m/z$ species, consistent with the greater mass of the Im anion compared to BF$_4$. These average $m/z$ values reflect the presence of large, complex organic species that may interact with exposed surfaces in an electrospray system differently compared to that of small atomic secondary ions, potentially leading to further fragmentation cascades, deposition, or electrochemical reactions. A subset of the detected peaks to 1000 amu with relative abundance greater than 0.1\% is listed in the Appendix Table \ref{fig: appendixquadplot}.

\subsection{Negative Secondary Species}
\subsubsection{Electrospray TOF-SIMS Spectra}
\begin{figure}[h!]
    \centering
    \begin{subfigure}[t]{0.78\textwidth}
        \centering
        \includegraphics[width=\textwidth]{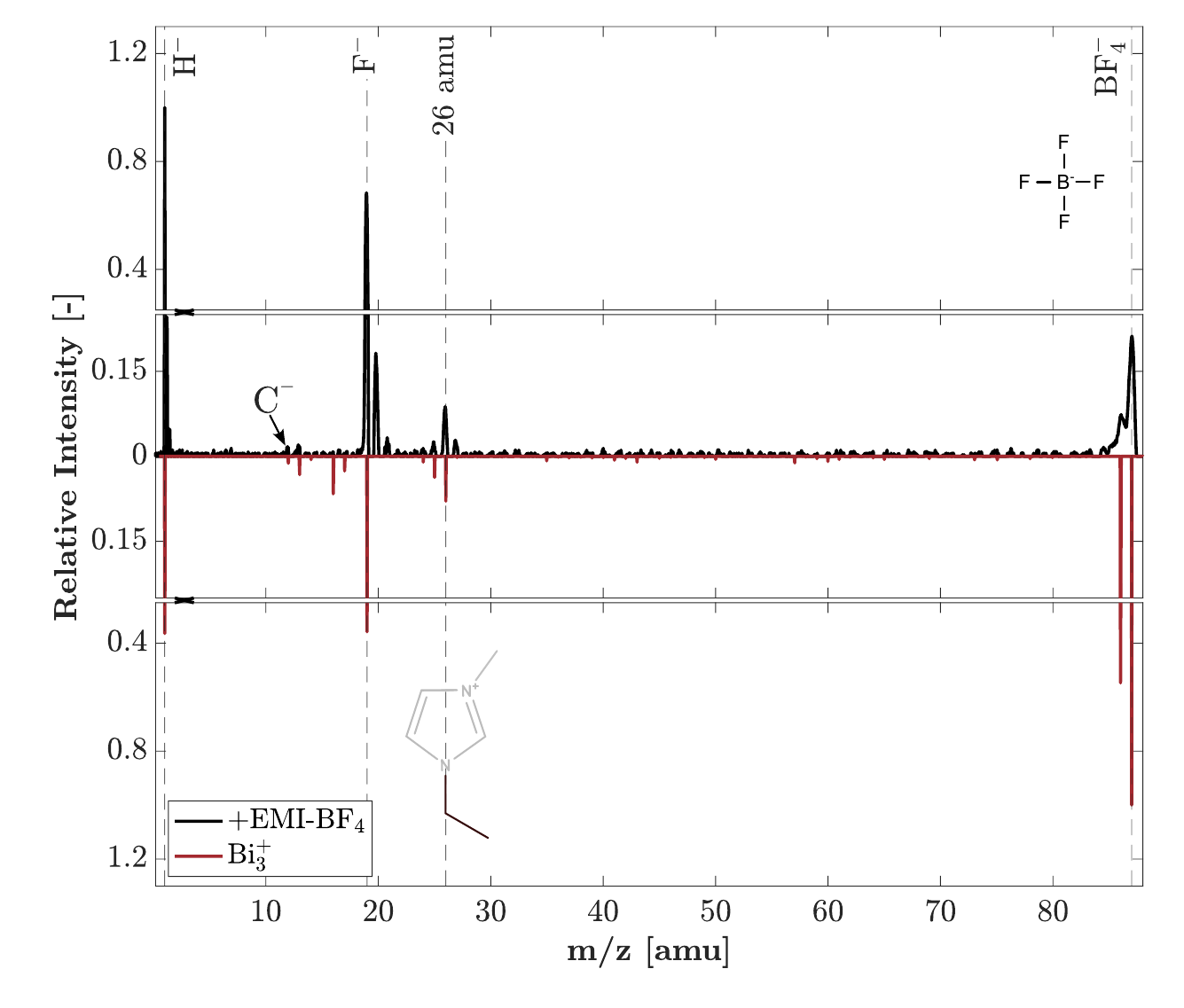}
        \caption{EMI-BF$_4$ thin film.}
        \label{fig:negEMIBF4}
    \end{subfigure}
    
    \vspace{0.75em}
    
    \begin{subfigure}[t]{0.78\textwidth}
        \centering
        \includegraphics[width=\textwidth]{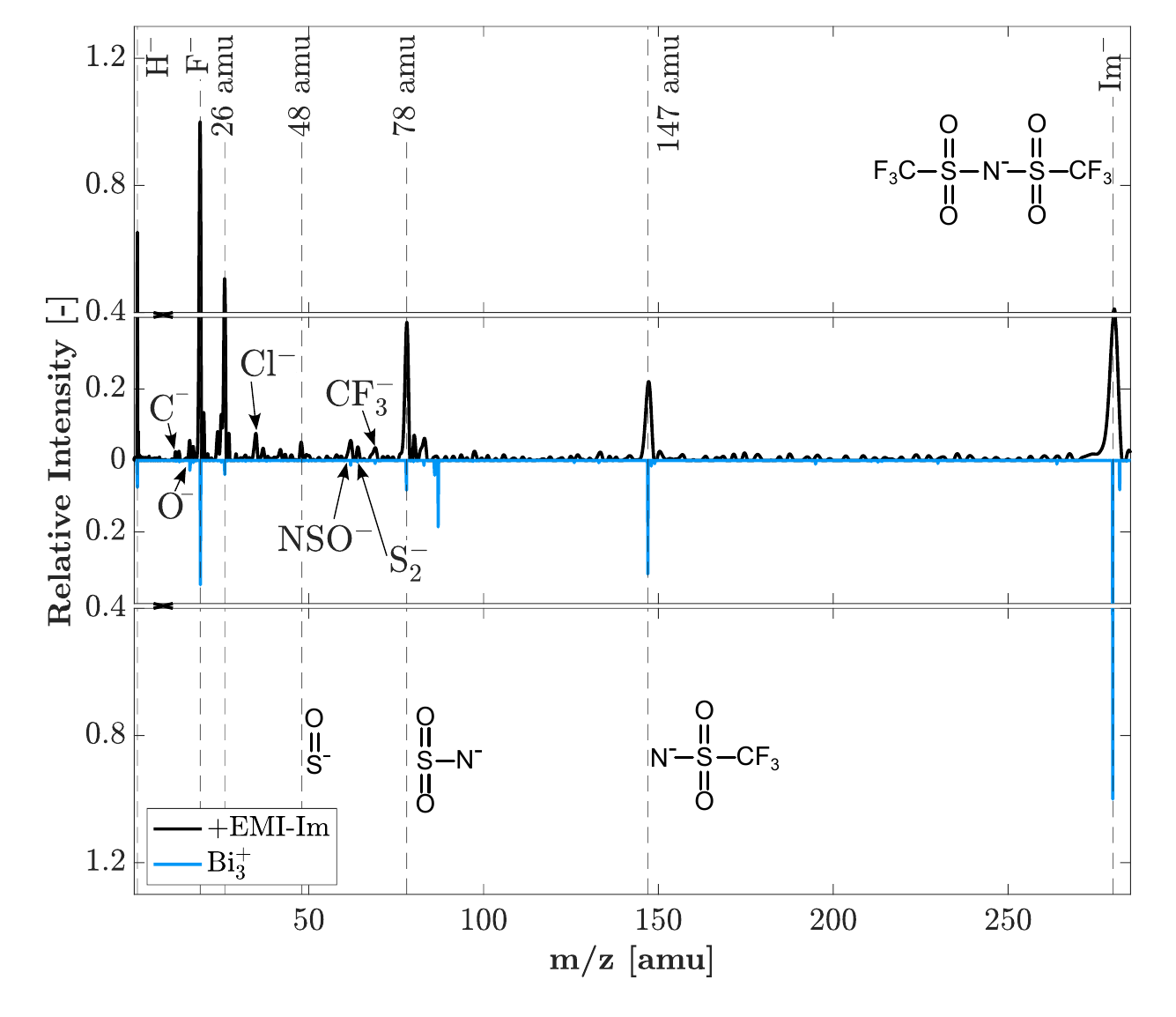}
        \caption{EMI-Im thin film.}
        \label{fig:negEMIIM}
    \end{subfigure}
    
    \caption{Negative secondary ion mass spectra of EMI-BF$_4$ and EMI-Im thin films exposed to (top of each subfigure) a 4 keV $E_\text{impact}$ positive electrospray plume and (bottom) a 25 keV Bi$_3^+$ primary ion beam in a commercial TOF-SIMS system. All spectra are normalized to the most intense peak. Structural diagrams of the Im$^-$ anion fragments are shown at the corresponding mass in the EMI-Im spectrum.}
    \label{fig:combinedNegSIMS}
\end{figure}

The negative secondary ion mass spectra generated from 4 keV maximum impact energy positive plume impacts with IL thin films are shown in Fig. \ref{fig:combinedNegSIMS}, with a table of detected peaks with relative abundances detailed in Table \ref{tab:negative species}. Again, also displayed in Fig. \ref{fig:combinedNegSIMS} are the corresponding negative secondary ion mass spectra from commercial SIMS of the IL thin film with a 25 keV Bi$_3^+$ primary beam. All spectra are individually normalized with respect to the most intense peak.

While the two ionic liquid films produce broadly similar positive ion fragmentation patterns, their negative secondary ion spectra diverge markedly, reflecting the greater chemical complexity of Im relative to BF$_4$. Even so, several common features are observed, including intense signals at $m/z$ = 1 (H$^-$) and 19 (F$^-$), as well as the respective parent anions at $m/z$ = 87 (BF$_4^-$) and 280 (Im$^-$). Both spectra also display characteristic hydrocarbon fragments, with peaks at $m/z$ = 12, 13 and in the 24–27 amu range, consistent with methyl- and ethyl-derived secondary ions.

The negative secondary ion spectrum of EMI–Im reveals prominent anion-derived fragments. A strong peak at $m/z$ = 147 corresponds to S–N bond cleavage, while another intense signal at $m/z$ = 78 arises from a two-bond scission that yields the NSO$_2^-$ ion. As seen in the raw TOF spectrum in the appendix (Fig. \ref{fig: appendixquadplot}), additional species are present between the 26 amu and 78 amu peaks. The first includes minor salt contamination from the substrate, with chlorine isotopes detected at $m/z$ = 35 and 37, consistent with previous work \cite{hofheins2025electrospray}. More definitively, a series of characteristic Im-derived fragments appear in this range, including SO$^-$ (48 amu), NSO$^-$ (62 amu), SO$_2^-$ (64 amu), and CF$_3^-$ (69 amu), consistent with established fragmentation pathways of the Im anion \cite{Bell2024}. In addition, a small peak at $m/z$ = 83 was detected with no obvious identification but can be seen in past SIMS work of EMI-Im films \cite{GUNSTER20083403}.

\subsubsection{IONTOF TOF-SIMS5 Spectra}
The commercial negative SIMS spectra broadly mirror the ESI-TOF SIMS results in terms of the dominant peaks, confirming consistency between the two techniques. In the EMI–BF$_4$ spectrum, additional structure is revealed in the low-mass region (26–87 amu), with BF$_2^-$ clearly resolved at $m/z$ = 48 and 49 amu. Beyond the EMI$^+$ monomer, larger cluster ions are observed, including a strong signal at the dimer, a small but detectable trimer peak, and even higher-order clusters extending up to $m/z$ = 730 amu. The EMI–Im spectra displays a similar correspondence to the ESI-TOF SIMS data in the major peaks, but with notable differences in the high-mass region: dimers and trimers are detected, including signal as high as $m/z$ = 1062 corresponding to the trimer, with families of fragments extending throughout the spectrum up to that mass. In addition, the higher resolution of the commercial instrument reveals a richer distribution of Im$^-$ fragments that were not as clearly distinguished in the ESI-TOF SIMS spectra, including F$_2^-$ (38 amu), S$^-$ (32 amu), S$_2$ON$^-$ (126 amu), CF$_3$SO$_2^-$ (133 amu), CF$_3$SO$_2$NSO$^-$ (195 amu), and CF$_3$SO$_2$NSOCF$_3^-$ (264 amu). Important to note is the detection of $m/z$ = 86, 87 amu in the negative EMI-Im spectrum, hinting at some level of EMI-BF$_4$ contamination. Overall, the commercial spectra mirror the intense features of the ESI-TOF SIMS data while providing increased resolution that enables identification of additional anion fragments as well as extended cluster families, ranging past the EMI–BF$_4$ dimer and up to the EMI–Im trimer. The average weighted mass-to-charge ratio of the EMI-BF$_4$ spectrum was 133 amu, compared to the EMI-Im at 209 amu, again highlighting the complex, relatively high $m/z$ ratio secondary ions formed from energetic ion bombardment.

\vspace{10pt}
\begin{table}
\centering
\caption{\label{tab:negative species} Negative secondary ion collection mode species identification, assuming singly charged species.}
\begin{tabular}{cccc}  
\hline \hline
Mass (amu) & Species Identification & \% Abundance EMI-BF$_4$  & \% Abundance EMI-Im \\ \hline
    1 & H$^-$ & 33.0 & 15.0 \\
    12 & C$^-$ & 1.2 & 0.2  \\
    13 & CH$^-$ & 2.1 & 0.6 \\
    16 & O$^-$ & --  & 1.2 \\
    17 & OH$^-$ & -- & 0.4 \\
    19 & F$^-$ & 28.7 & 23.3 \\
    24 & C$_2^-$ & 1.2 & 1.9 \\
    25 & C$_2$H$^-$ & 2.0 & 3.8 \\
    26 & C$_2$H$_2^-$ & 5.7 & 10.4 \\
    35 & $^{35}$Cl$^-$ & -- & 2.1 \\
    37 & $^{37}$Cl$^-$ & -- & 0.8 \\
    48 & SO$^-$ & -- & 1.0 \\
    62 & NSO$^-$ & --  &  0.7 \\
    64 & SO$_2^-$ & --  &  0.7 \\
    69 & CF$_3^-$ & -- & 0.4 \\
    78 & NSO$_2^-$ & -- & 3.7 \\
    83 &  & -- & 1.7 \\
    86 & $^{10}$BF$_4^{-}$ & 8.2 & -- \\ 
    87 & $^{11}$BF$_4^{-}$ & 15.6 & --\\
    147 & CF$_3$NSO$_2^-$ &  -- & 6.3 \\
    280 & Im$^{-}$ & --  & 24.9 \\
    284 & BF$_4^-$(EMI-BF$_4$) & 0.5 & --\\
    cont. $m/z$ & & 1.7 & 1.3 \\
    \hline \hline
\end{tabular}
\end{table}


\subsection{Post Processing}
\label{sec: opticalprof}
To examine the target wafers post-firing by the electrospray TOF-SIMS diagnostic, a Zygo NewView Optical Profilometer at the Cornell NanoScale Facility was employed. This analytical tool provides non-destructive, relative surface topology measurements with resolution down to 0.1 nm using white light interferometry. This tool was therefore used to characterize sputter patterns, depth, and thin film thickness of the target substrates.

\begin{figure}[h!]
    \centering
    \includegraphics[width=1\textwidth]{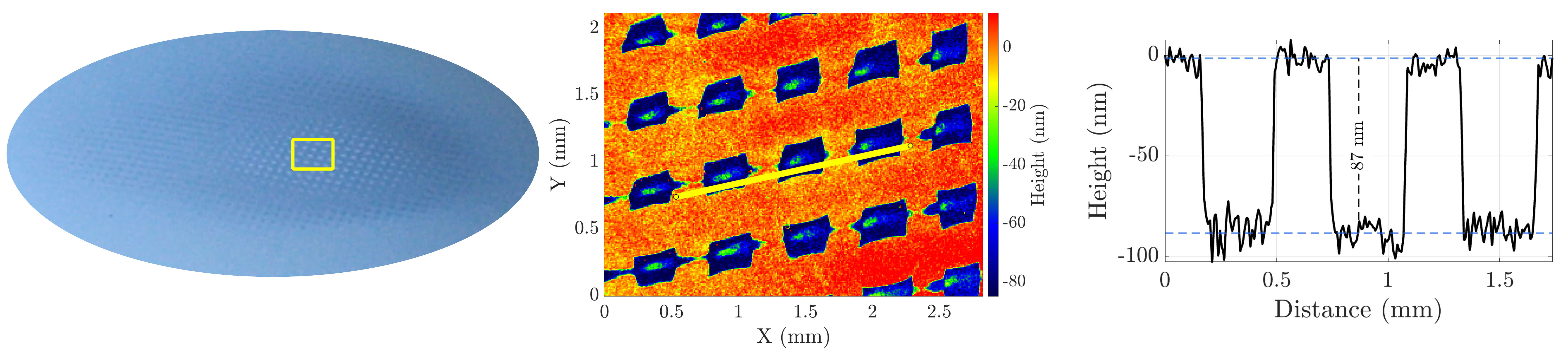}
    \caption{Optical Profilometer 2D surface map (middle) of the exposed IL thin film post-SIMS analysis of the center the beam spot (left), with a corresponding 1D line horizontal cross-section line profile (right). The grid pattern forms due to electrical and physical blocking by the secondary ion acceleration grid in front of the target.}
    \label{fig: profilometer}
\end{figure}

The positive EMI-BF$_4$ plume was exposed to the IL thin film target for 7 minutes in order acquire the negative TOF-SIMS spectrum, after which the source was maintained in negative plume mode for 173 minutes to ensure stable firing conditions for the profilometry sputter-rate analysis. Post-removal from the vacuum chamber testing facility, a clear sputtered beam spot is evident on the target film wafer. Marked by a grid pattern formed from the physical and electrical blocking of the semi-transparent stainless steel acceleration grid mounted to the target, this allowed for approximations of the sputter rate of the film through analysis via optical profilometry. Relative surface topology at 2.5X magnification of the sputtered pattern are shown in a 2 mm x 2.8 mm field-of-view are in Fig. \ref{fig: profilometer}.

Given that the optical profilometer gives relative differences in surface height topologies, estimates of total sputter depth are made across horizontal cross-sections through the grid pattern of each section of the beam spot. The average distance in height from the unsputtered bulk film to the exposed area was determined to be 87 $\pm$ 3 nm, equivalent to a 0.48 nm/min sputter rate in the center of the sputtered spot, averaged across 50 cross sections.

By leveraging the optical profilometer's ability to quantify step heights, an estimate of total film thickness was made as well. To do this, all deposited material was removed with a fine-tipped instrument near the sputtered spot. Measuring the step height of 50 random cross sections between the bottom of the swipe and the bulk film resulted in a total thickness of 130 nm $\pm$ 11 nm.

While silver was chosen as the ionic liquid (IL) film substrate due to improved wetting compared to other targets such as gold, visual discoloration of the target and the removal of both ionic liquid and silver during surface cleaning indicate chemical interaction and partial dissolution of silver into the IL. Although 100 nm of silver was initially deposited onto the silicon wafer and subsequently spin-coated with 3 mL of an ionic-liquid–methanol solution, the resulting measured film thickness of 130 nm does not necessarily correspond to a net 30 nm ionic liquid layer. Prior ionic-liquid thin-film studies have shown that a 1:20 1-butyl-3-methylimidazolium bis(trifluoromethylsulfonyl)imide (BMI-Im) to methanol solution yields an approximately 120 nm thin film \cite{boning_ILthinfilm}. It is therefore hypothesized that dissolution of silver into the bulk ionic-liquid film leads to a composite silver–IL layer with an effective thickness of approximately 130 nm.

This hypothesis is supported by the consistent ESI-TOF-SIMS spectra observed over the three-hour firing duration, which lacked the silver isotope and silver cluster ion peaks characteristic of bare silver targets \cite{hofheins2025electrospray}. Given the likelihood of silver dissolution into the bulk ionic-liquid film, the measured film thickness and the inferred sputter depth and sputter rate should be regarded as representative only of this composite silver–IL target under the present experimental conditions, with additional uncertainty arising from processes such as redeposition or re-adsorption of sputtered neutrals. Dedicated measurements on bare metallic substrates and independently prepared, well-characterized ionic-liquid thin films are required to robustly quantify absolute sputter depths and sputter rates.

\setlength{\intextsep}{1pt}
\vspace{0pt}

\section{Conclusion}

Secondary ion mass spectrometry analysis of ionic liquid thin films was applied	through a laboratory diagnostic (ESI TOF-SIMS) and analytical grade SIMS system (IONTOF5) in order to determine mass-to-charge ratio and chemical composition of electrospray-relevant secondary species. Collisions of the ionic liquid thin film both with a negative electrospray plume and a cluster bismuth primary ion beam revealed mirrored positive secondary ion mass spectra for EMI-BF4 and EMI-Im thin films, mirroring the shared cation. The spectra are dominated by EMI$^+$ ($m/z$ = 111), with sequential side-chain stripping (neutral methyl losses) producing fragments at $m/z$ = 96, 83, and 68, and lower-mass ions at $m/z$ = 55, 42, 28, and 12 arising from imidazolium ring cleavage. While fragments of the BF$_4^-$ anion were evident in the EMI-BF$_4$ diagnostic positive spectrum, no obvious analogous Im$^-$ fragments were detected in the positive EMI-Im spectrum. In terms of higher-$m/z$ species, the laboratory diagnostic only distinctly detected the positive dimer for both films, whereas the higher-resolution commercial spectra resolved additional peaks of small relative intensity above the trimer, shifting the average $\langle m/z \rangle_{\mathrm{EMI\text{-}BF_4}} = 123~\text{amu}$ and $\langle m/z \rangle_{\mathrm{EMI\text{-}Im}} = 138~\text{amu}$ above that of the monomer.

In comparison, the negative secondary ion mass spectra from the two IL thin films diverged markedly correlated to the difference in chemical composition of the BF$_4$ and Im anions. Both ESI TOF-SIMS spectra contained the monomer anions as well as strong peaks for hydride and fluorine, but EMI-Im produced numerous anion fragments, dominated by single-bond cleavages. The laboratory diagnostic revealed negative dimers detected for EMI-BF$_4$, but not clearly for EMI-Im. Again, the commercial data were consistent with the laboratory diagnostic but revealed higher-$m/z$ species with greater resolution. Owing to the larger Im anion, the average $m/z$ of the negative EMI-Im spectrum was 209 amu, about 80 amu greater than that of EMI-BF$_4$. 

Finally, complementary optical profilometry of a bombarded EMI-BF$_4$ thin film indicated possible dissolution of the silver substrate into the ionic liquid layer, but nonetheless provided rough estimates of film thickness, sputter depth, and sputter rate. 

These results provide direct, species-resolved experimental measurements of the secondary species previously reported in vacuum-facility and plume–surface interaction studies by Uchizono et al. \cite{uchizono_role_2021, uchizono_facility_2022, uchizono2022positive} and Geiger et al. \cite{geiger2025secondary}, and directly link these observations to the formation pathways associated with secondary species identified during electrospray plume–extractor impacts by Kerber et al. \cite{kerber2025characterizationpt1, kerber2025characterization}. More broadly, these experimentally measured secondary-ion populations are consistent with, and provide experimental grounding for, foundational numerical studies of ionic-liquid plume impacts using classical molecular dynamics by Tahsin et al. \cite{tahsin_reactive_2025} and Bendimerad et al. \cite{bendimerad2022molecular}, as well as quantum-based simulations by Laws et al. \cite{laws2025emiBF4surfaceJAP}. Therefore, these results connect prior secondary species literature by bridging experimental observations with existing numerical studies to provide a consolidated experimental basis for interpreting secondary species emission in ionic-liquid electrospray systems.

In summary, the secondary ion spectra demonstrate that plume--surface interactions generate not only atomic secondary ions but also a broad continuum of molecular fragments and clusters extending to high $m/z$, underscoring the intrinsic chemical complexity of secondary ion formation. The resulting secondary ions can interact with electric fields, backstream to the emitter, erode the extractor electrode, and facilitate electrochemical reactions. Over long durations, such backstreaming species contribute to emitter degradation and reduced thruster lifetime.  In ground-based performance evaluation of ionic liquid electrospray thrusters, the plume eventually impacts a surface – whether that surface is a dedicated beam target, a diagnostic face, or chamber walls. The aggregate of these secondary ions depositing on surfaces, like the top-side of the extractor electrode, over the course of  lifetime tests of electrospray thrusters contribute to premature extractor and emitter degradation and thus system failure. Therefore, the significance of the ionic liquid secondary ions formed from plume-ionic liquid surface interactions are paramount to understanding electrospray lifetime limiting processes, through the lens of both intrinsic thruster operation as well as facility effects. 

\ack{}

\funding{This work was supported by NASA Space Technology Graduate Research Opportunity Fellowship (80NSSCK23K1212). This work was performed in part at the Cornell NanoScale Facility, a member of the National Nanotechnology Coordinated Infrastructure (NNCI), which is supported by the National Science Foundation (Grant NNCI-2025233).}

\roles{
Giuliana Hofheins: Conceptualization, Methodology, Investigation, Data Curation, Formal Analysis, Visualization, Writing – Original Draft.
Aleksandra Biedron: Investigation, Validation, Resources.
Elaine Petro: Conceptualization, Supervision, Funding Acquisition, Writing – Review \& Editing.}

\data{The data that support the findings of this study are available
from the corresponding author upon request}

\suppdata{
\newpage
\vspace{0.5\baselineskip}  
\small
\setlength{\tabcolsep}{5pt}        
\renewcommand{\arraystretch}{0.85} 

\begin{longtable}{@{}rrrrrrrr@{}}
\caption{Detected peaks with $\geq$ 0.1\% relative abundance from IONTOF TOF-SIMS---5 system, in both polarities (positive ($+$) and negative ($-$)) from EMI-Im and EMI-BF$_4$ thin films. Monomer and dimer peaks are bolded.}\label{tab:ualbany-data}\\
\hline\hline
\multicolumn{4}{c}{\textbf{EMI-Im}} & \multicolumn{4}{c}{\textbf{EMI-BF$_4$}} \\
\cline{1-4}\cline{5-8}
\textbf{$+$} [amu] & Rel. Abun. [\%] & \textbf{$-$} [amu] & Rel. Abun. [\%] & \textbf{$+$} [amu] & Rel. Abun. [\%] & \textbf{$-$} [amu] & Rel. Abun. [\%] \\
\hline
\endfirsthead

\caption[]{IONTOF5 Peaks (continued)}\\
\hline\hline
\multicolumn{4}{c}{\textbf{EMI-Im}} & \multicolumn{4}{c}{\textbf{EMI-BF$_4$}} \\
\cline{1-4}\cline{5-8}
$+$ & Rel. Abun. [\%] & $-$ & Rel. Abun. [\%] & $+$ & Rel. Abun. [\%] & $-$ & Rel. Abun. [\%] \\
\hline
\endhead

\hline\multicolumn{8}{r}{\small\itshape Continued on next page}\\
\endfoot

\hline\hline
\endlastfoot

        1.008 & 0.7 & 1.008 & 3.2 & 1.007 & 0.3 & 1.008 & 13.5  \\ 
        11.008 & 0.2 & 12.001 & 0.1 & 10.012 & 0.1 & 12.001 & 0.5  \\ 
        12.000 & 0.2 & 13.009 & 0.2 & 11.009 & 0.2 & 13.009 & 1.2  \\ 
        13.007 & 0.2 & 15.994 & 1.2 & 13.006 & 0.1 & 15.997 & 2.5 \\ 
        14.015 & 0.4 & 17.002 & 0.2 & 14.015 & 0.2 & 17.005 & 1.0  \\ 
        15.024 & 1.3 & 19.000 & 14.8 & 15.023 & 1.1 & 18.999 & 13.3  \\ 
        26.015 & 0.7 & 25.008 & 0.2 & 22.991 & 0.1 & 24.001 & 0.4  \\ 
        27.022 & 2.1 & 26.005 & 1.7 & 26.013 & 0.4 & 25.008 & 1.4  \\ 
        28.020 & 2.2 & 31.974 & 0.1 & 27.024 & 2.3 & 26.004 & 2.9 \\ 
        29.039 & 2.7 & 61.971 & 0.6 & 28.019 & 1.2 & 34.971 & 0.3 \\ 
        30.037 & 0.3 & 63.964 & 0.2 & 29.038 & 1.9 & 43.019 & 0.4 \\ 
        30.997 & 0.2 & 68.998 & 0.4 & 30.035 & 0.2 & 57.036 & 0.4  \\ 
        33.015 & 0.1 & 77.970 & 3.6 & 31.019 & 1.0 & 59.985 & 0.3  \\ 
        38.014 & 0.1 & 79.966 & 0.2 & 38.013 & 0.1 & \textbf{86.013} & \textbf{20.3}  \\ 
        39.025 & 0.4 & 82.965 & 0.6 & 39.023 & 1.5 & \textbf{87.004} & \textbf{37.1}  \\ 
        40.018 & 0.4 & 86.014 & 1.8 & 40.032 & 0.2 & 94.011 & 0.5  \\ 
        41.028 & 0.7 & 87.011 & 7.9 & 41.038 & 3.0 & 120.942 & 0.9  \\ 
        42.038 & 4.3 & 96.970 & 0.1 & 42.037 & 2.9 & 123.942 & 0.3  \\ 
        43.041 & 0.2 & 125.940 & 0.3 & 43.056 & 0.8 & 279.913 & 2.7  \\ 
        44.052 & 0.2 & 132.965 & 0.3 & 44.051 & 0.1 & 285.112 & 0.3  \\ 
        44.976 & 0.1 & 146.975 & 13.5 & 45.032 & 0.2 & ~ & ~  \\ 
        51.024 & 0.1 & 147.975 & 0.7 & 50.012 & 0.1 & ~ & ~  \\ 
        52.019 & 0.3 & 148.964 & 0.4 & 51.023 & 0.1 & ~ & ~ \\ 
        53.027 & 0.1 & 194.938 & 0.4 & 52.018 & 0.1 & ~  & ~ \\ 
        54.037 & 1.0 & 210.931 & 0.1 & 53.040 & 0.3 & ~ & ~ \\ 
        55.042 & 0.6 & 212.772 & 0.2 & 54.036 & 0.8 & ~ & ~ \\ 
        56.052 & 1.0 & 213.939 & 0.1 & 55.055 & 0.6 & ~ & ~ \\ 
        57.058 & 0.1 & 229.934 & 0.4 & 56.051 & 1.0 & ~ & ~ \\ 
        58.067 & 0.1 & 263.939 & 0.4 & 57.034 & 0.3 & ~& ~ \\ 
        60.024 & 0.1 & \textbf{279.938} & \textbf{42.6} & 58.044 & 0.1 & ~ & ~ \\ 
        66.034 & 0.1 & 281.927 & 3.5 & 59.053 & 0.6 & ~  & ~ \\ 
        67.041 & 0.1 & & ~ & 61.021 & 0.1 & ~ & ~  \\ 
        68.050 & 0.4 & & ~ & 62.032 & 0.1 & ~ & ~ ~ \\ 
        69.046 & 0.4 &  & ~ & 63.007 & 0.1 & ~ &  ~ \\ 
        70.070 & 0.1 &  & ~ & 64.981 & 0.1 & ~ &  ~ \\ 
        73.052 & 0.1 & ~ & ~ & 66.034 & 0.1 & ~ &  ~ \\ 
        80.048 & 0.1 & ~ & ~ & 67.055 & 0.1 & ~ &  ~ \\ 
        81.046 & 1.0 & ~ & ~ & 68.054 & 0.3 & ~ & ~~ \\ 
        82.056 & 3.6 & ~ & ~ & 69.046 & 0.3 & ~ & ~ \\ 
        83.061 & 4.6 & ~ & ~ & 70.070 & 0.1 &  & ~ \\ 
        84.183 & 0.2 & ~ & ~ & 73.057 & 2.2  & ~ & ~ \\ 
        94.054 & 0.1 & ~ & ~ & 74.052 & 0.2  & ~ & ~ \\ 
        95.058 & 1.3 & ~ & ~ & 75.028 & 0.1  & ~ & ~ \\ 
        96.068 & 0.9 & ~ & ~ & 76.048 & 0.1  & ~ & ~ \\ 
        97.078 & 0.5 & ~ & ~ & 77.037 & 0.1  & ~ & ~ \\ 
        101.052 & 0.1 & ~ & ~ & 79.056 & 0.1 & ~  & ~ \\ 
        107.058 & 0.2 & ~ & ~ & 80.053 & 0.1 & ~ & ~ \\ 
        108.067 & 0.1 & ~ & ~ & 81.046 & 0.8  & ~ & ~ \\ 
        109.080 & 6.9 & ~ & ~ & 82.056 & 2.9  & ~ & ~ \\ 
        110.092 & 5.3 & ~ & ~ & 83.061 & 4.4  & ~ & ~ \\ 
        \textbf{111.095} & \textbf{45.2} & ~ & ~ & 84.178 & 0.2  & ~ & ~ \\ 
        112.090 & 2.1 & ~ & ~ & 91.055 & 0.1  & ~ & ~ \\ 
        121.072 & 0.1 & ~ & ~ & 94.054 & 0.1  & ~ & ~ \\ 
        122.084 & 0.1 & ~ & ~ & 95.065 & 1.1  & ~ & ~ \\ 
        123.093 & 0.4 & ~ & ~ & 96.069 & 0.9  & ~ & ~ \\ 
        124.100 & 0.2 & ~ & ~ & 97.078 & 0.6  & ~ & ~ \\ 
        125.111 & 1.0 & ~ & ~ & 101.053 & 0.1  & ~ & ~ \\ 
        126.112 & 0.1 & ~ & ~ & 103.042 & 0.2  & ~ & ~ \\ 
        127.083 & 0.1 & ~ & ~ & 107.060 & 0.2  & ~ & ~ \\ 
        129.085 & 0.5 & ~ & ~ & 108.068 & 0.1  & ~ & ~ \\ 
        130.088 & 0.1 & ~ & ~ & 109.081 & 5.7  & ~ & ~ \\ 
        135.091 & 0.2 & ~ & ~ & 110.093 & 3.6  & ~ & ~ \\ 
        136.089 & 0.1 & ~ & ~ & 110.502 & 0.1  & ~ & ~ \\ 
        137.105 & 0.1 & ~ & ~ & \textbf{111.097} & \textbf{43.5}  & ~ & ~ \\ 
        139.126 & 0.1 & ~ & ~ & 112.098 & 4.0  & ~ & ~ \\ 
        147.067 & 0.1 & ~ & ~ & 113.104 & 0.3  & ~ & ~ \\ 
        216.983 & 0.1 & ~ & ~ & 115.071 & 0.1  & ~ & ~ \\ 
        218.984 & 0.1 & ~ & ~ & 121.081 & 0.1  & ~ & ~ \\ 
        308.228 & 0.3 & ~ & ~ & 122.086 & 0.1  & ~ & ~ \\ 
        309.231 & 1.3 & ~ & ~ & 123.095 & 0.5  & ~ & ~ \\ 
        310.226 & 0.2 & ~ & ~ & 124.102 & 0.2  & ~ & ~ \\ 
        327.073 & 0.4 & ~ & ~ & 125.113 & 1.3  & ~ & ~ \\ 
        328.074 & 0.1 & ~ & ~ & 126.114 & 0.1  & ~ & ~ \\ 
        329.077 & 0.3 & ~ & ~ & 129.087 & 0.4  & ~ & ~ \\ 
        ~ & ~ & ~ & ~ & 130.097 & 0.1 & ~ & ~  \\ 
        ~ & ~ & ~ & ~ & 131.035 & 0.1 & ~ &  ~ \\ 
        ~ & ~ & ~ & ~ & 133.053 & 0.1 & ~ & ~ \\ 
        ~ & ~ & ~ & ~ & 135.094 & 0.2 & ~ & ~ \\ 
        ~ & ~ & ~ & ~ & 136.099 & 0.1 & ~  & ~ \\ 
        ~ & ~ & ~ & ~ & 137.115 & 0.2 & ~ & ~ \\ 
        ~ & ~ & ~ & ~ & 139.129 & 0.1 & ~ & ~ \\ 
        ~ & ~ & ~ & ~ & 143.100 & 0.1  & ~ & ~ \\ 
        ~ & ~ & ~ & ~ & 147.075 & 0.5  & ~ & ~ \\ 
        ~ & ~ & ~ & ~ & 148.079 & 0.1  & ~ & ~ \\ 
        ~ & ~ & ~ & ~ & 207.035 & 0.1  & ~ & ~ \\ 
        ~ & ~ & ~ & ~ & \textbf{308.229} & 0.1  & ~ & ~ \\ 
        ~ & ~ & ~ & ~ & \textbf{309.233} & 0.3  & ~ & ~ \\ 
        ~ & ~ & ~ & ~ & 310.228 & 0.1  & ~ & ~ \\ 
        ~ & ~ & ~ & ~ & 481.501 & 0.1  & ~ & ~ \\ 
        ~ & ~ & ~ & ~ & 483.530 & 0.3  & ~ & ~ \\ 
        ~ & ~ & ~ & ~ & 484.533 & 0.1  & ~ & ~ \\ \hline\hline

\end{longtable}

}

\appendix 
\section{Raw ESI TOF-SIMS Spectra}

\begin{figure}[htbp]
    \centering
    
    \begin{subfigure}{0.495\textwidth}
        \centering
        \includegraphics[width=\linewidth]{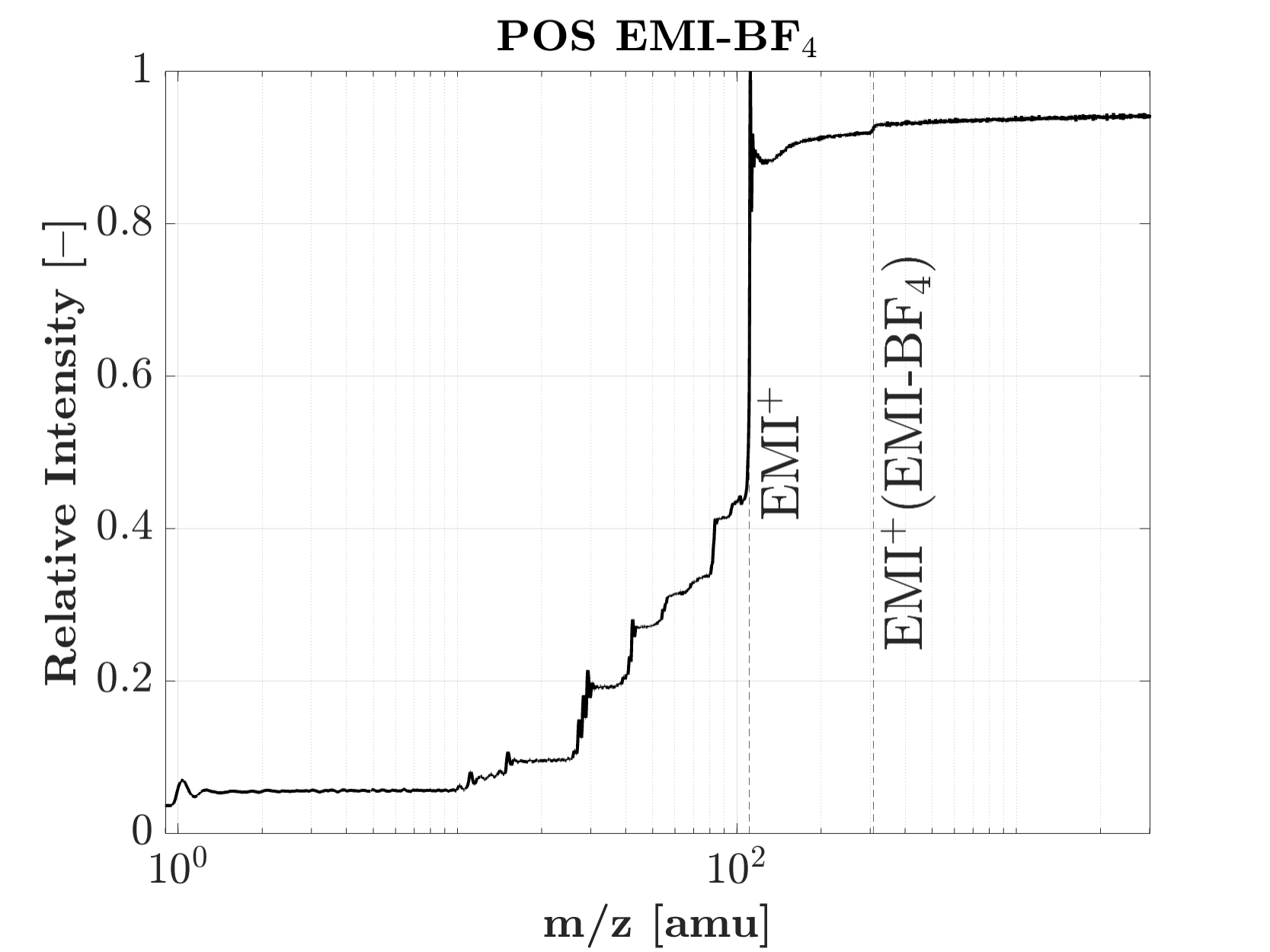}
    \end{subfigure}
    \hfill
    \begin{subfigure}{0.495\textwidth}
        \centering
        \includegraphics[width=\linewidth]{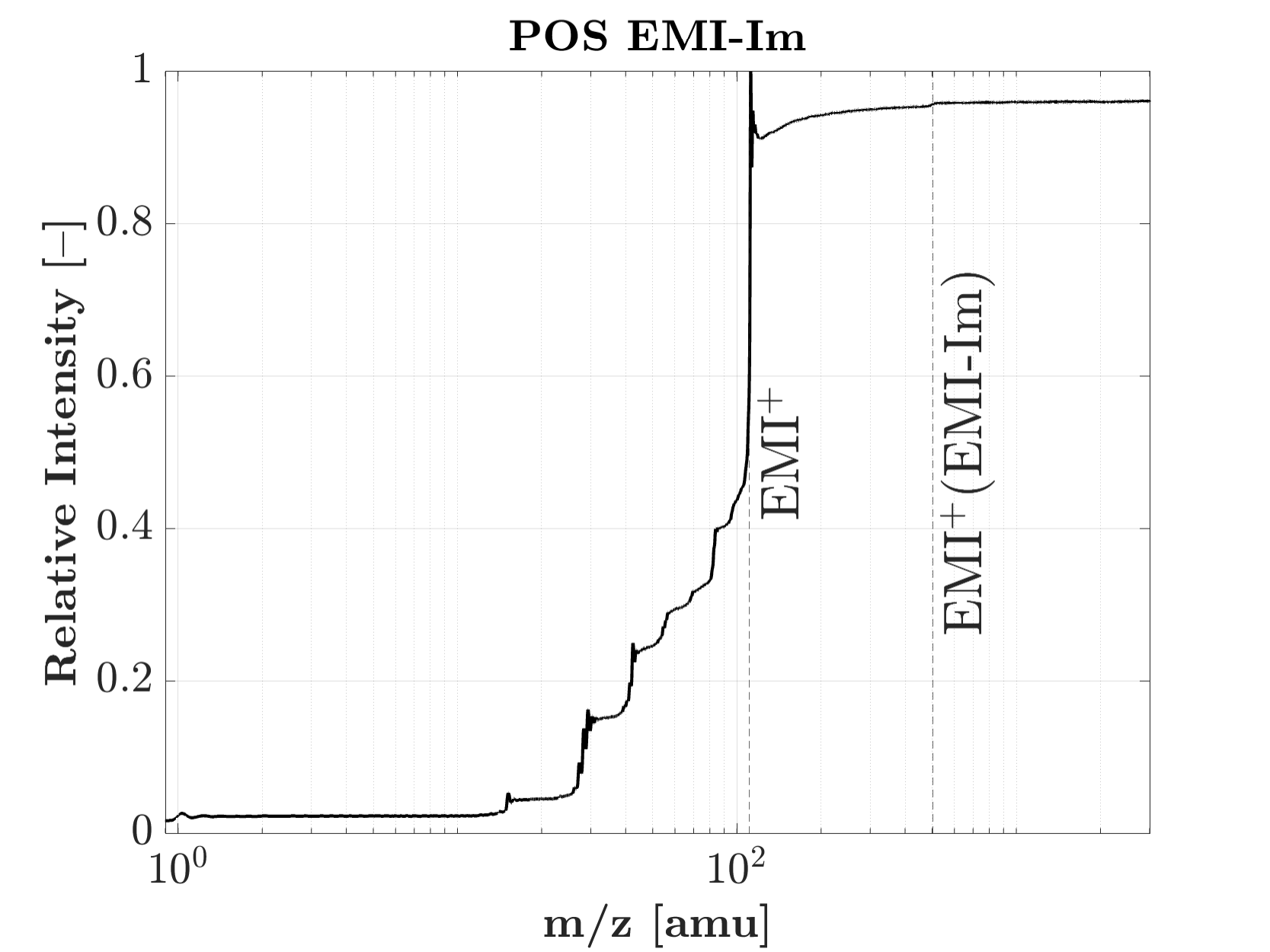}
    \end{subfigure}
    
    \vskip\baselineskip
    
    \begin{subfigure}{0.495\textwidth}
        \centering
        \includegraphics[width=\linewidth]{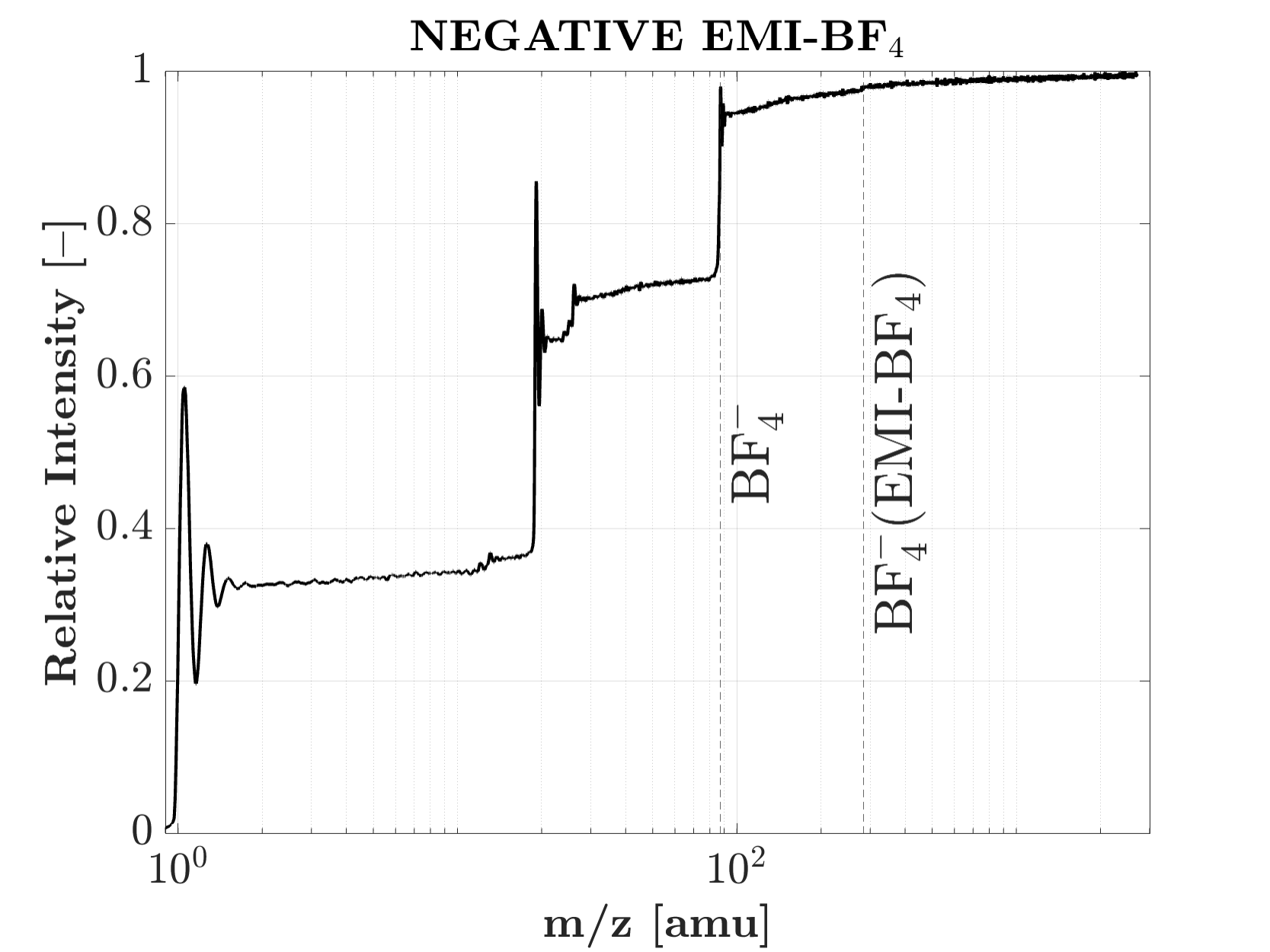}
    \end{subfigure}
    \hfill
    \begin{subfigure}{0.495\textwidth}
        \centering
        \includegraphics[width=\linewidth]{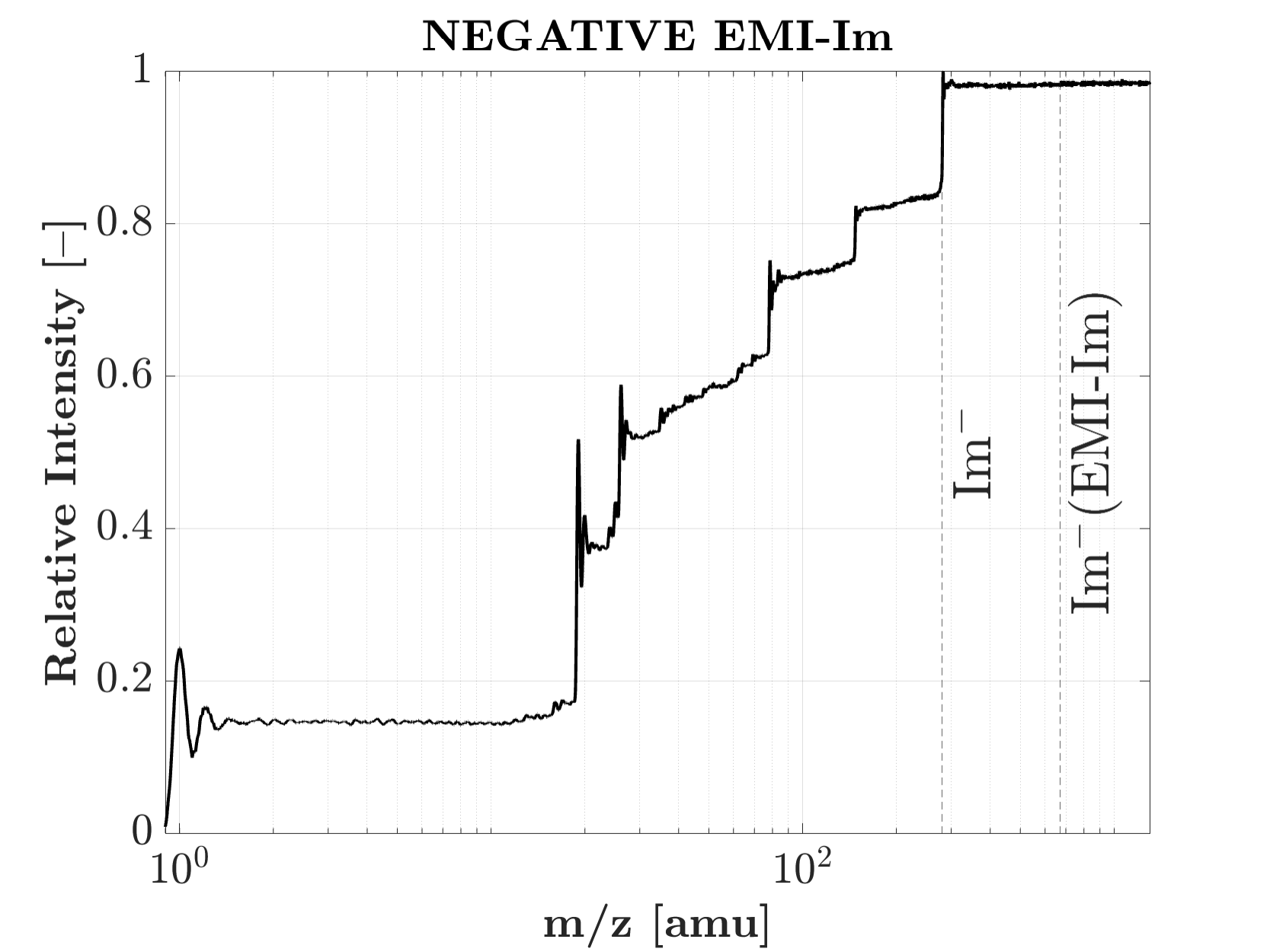}
    \end{subfigure}
    
    \caption{Raw time-of-flight curves corresponding to derivative plots in Figures \ref{fig:combinedPosSIMS} and \ref{fig:combinedNegSIMS}.}
    \label{fig: appendixquadplot}
\end{figure}

\section{Log-Scale IONTOF 5 Spectra}
\begin{figure}[htbp]
    \centering
    \begin{subfigure}{0.495\textwidth}
        \centering
        \includegraphics[width=\linewidth]{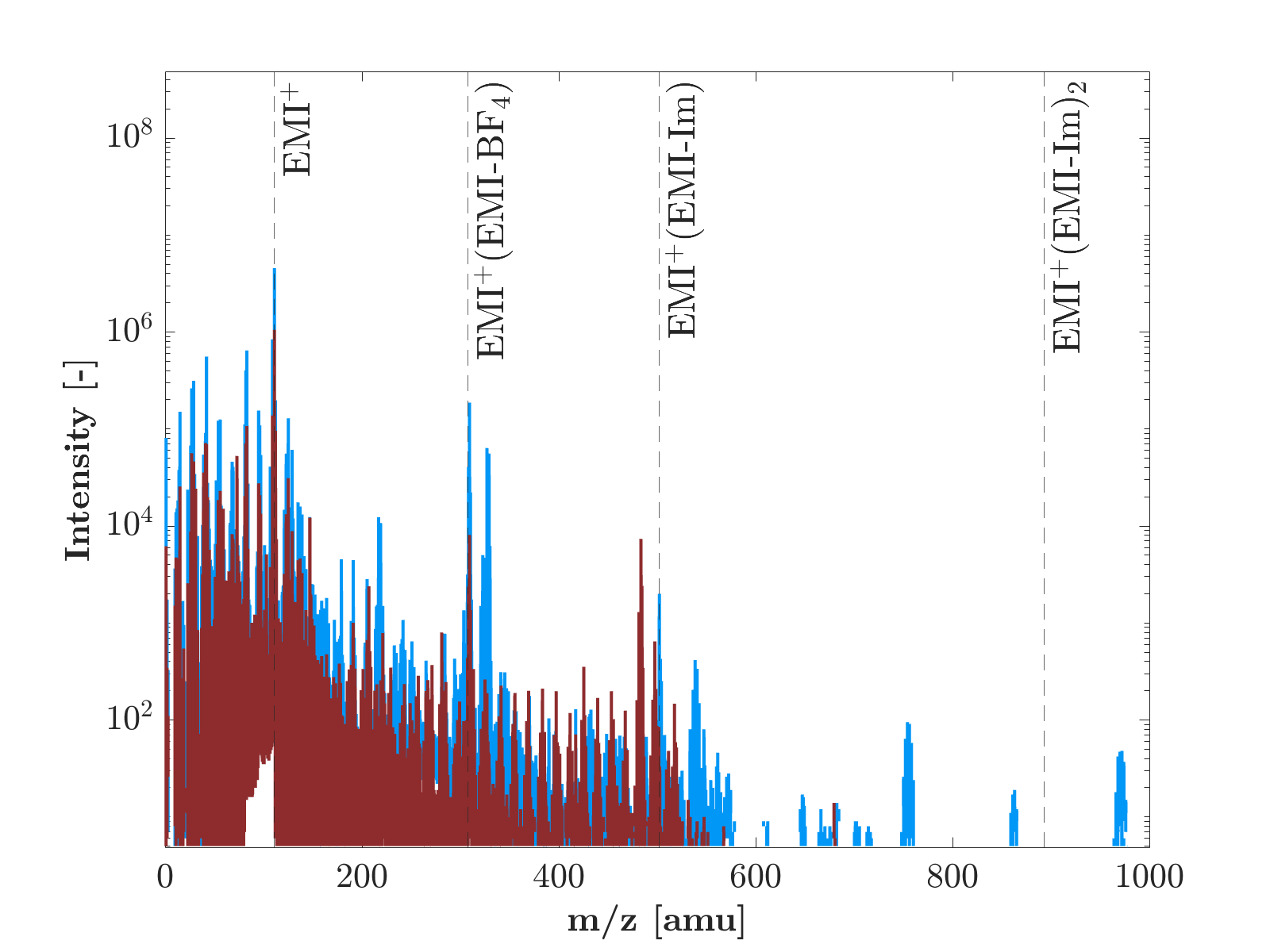}
    \end{subfigure}
    \hfill
    \begin{subfigure}{0.495\textwidth}
        \centering
        \includegraphics[width=\linewidth]{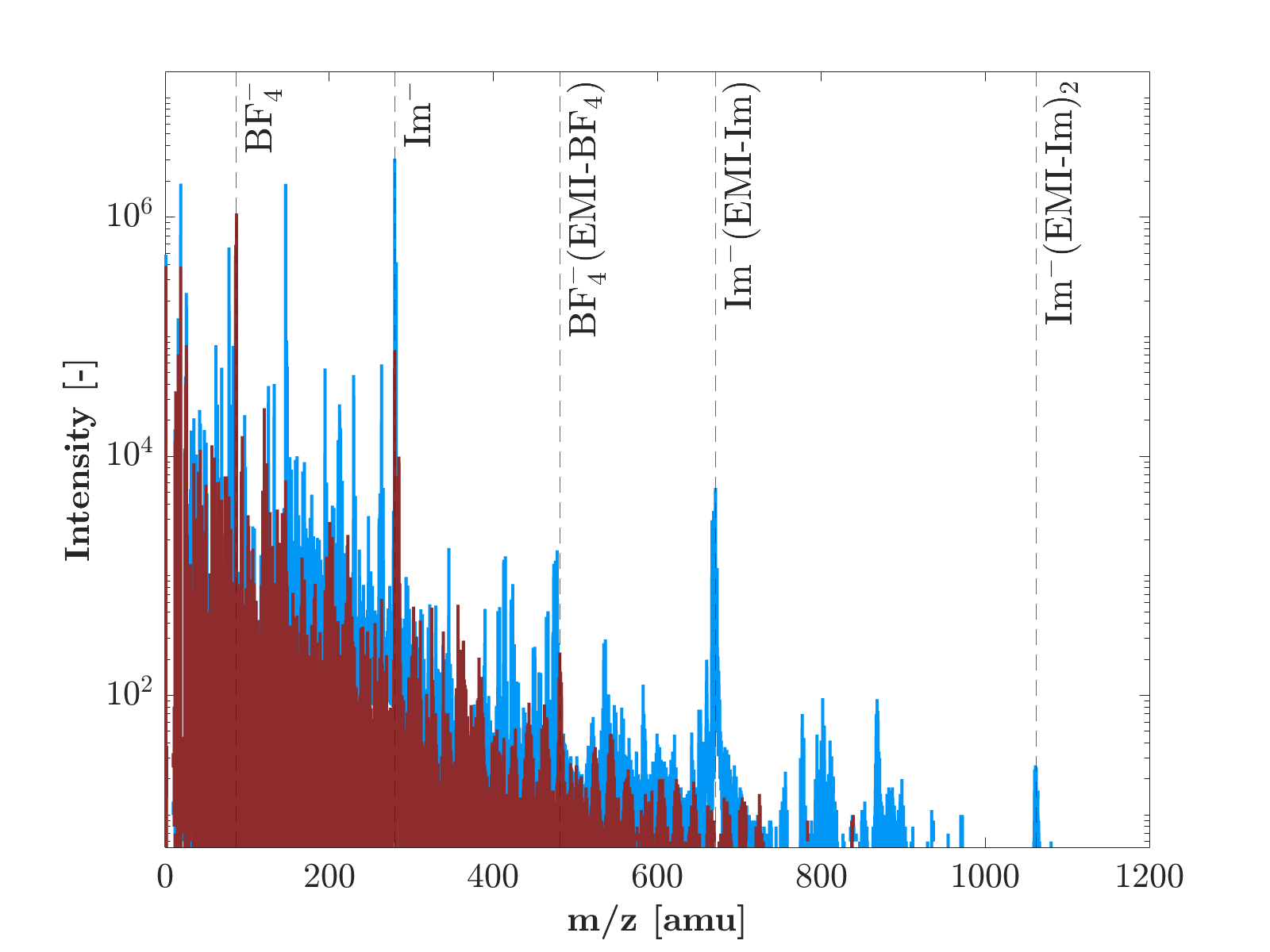}
    \end{subfigure}
    
    \caption{Log-scale SIMS spectra from the Bi$_3^+$ commercial system corresponding to $<$ 111 amu plots in Figures \ref{fig:combinedPosSIMS} and \ref{fig:combinedNegSIMS} }
    \label{fig: appendixlogscale}
\end{figure}

\newpage
\printbibliography

\end{document}